\documentclass[a4paper,12pt]{article}
\usepackage[utf8]{inputenc}
\usepackage[english]{babel}
\usepackage{amsmath, amssymb,graphicx,bm}
\usepackage{braket}
\usepackage{caption}
\usepackage{subcaption}
\usepackage{ulem}
\usepackage{multicol}

\pdfoutput=1
\usepackage{jheppubm}
\newcommand{\be}{\begin{equation}}
\newcommand{\ee}{\end{equation}}
\newcommand{\bea}{\begin{eqnarray}}
\newcommand{\eea}{\end{eqnarray}}
\newcommand{\nn}{\nonumber}

\newcommand{\fb}{\mathfrak{b}}

\newcommand{\fg}{\mathfrak{g}}

\newcommand{\fp}{\mathfrak{g}}

\newcommand{\e}{\mathrm{e}}

\definecolor{darkraspberry}{rgb}{0.53, 0.15, 0.34}

\definecolor{darkblue}{rgb}{0., 0, 1}

\definecolor{dgreen}{rgb}{0.,0.6,0.}

\title{Direct photons emission rate and electric conductivity  in twice anisotropic QGP holographic model with first-order phase transition}

\author{Irina Ya. Aref'eva$^a$, Alexey Ermakov$^a$ and Pavel Slepov$^a$}

\affiliation{$^a$Steklov Mathematical Institute, Russian Academy of
  Sciences,\\ Gubkina str. 8, 119991, Moscow, Russia}

\emailAdd{arefeva@mi-ras.ru}
\emailAdd{ermakov.av15@physics.msu.ru}
\emailAdd{slepov@mi-ras.ru}

\abstract{The electric conductivity and direct photons emission rate are considered in the holographic theory with two types of  anisotropy. The electric conductivity is derived in two different ways, and their equivalence for the twice anisotropic theory is shown.  Numerical calculations of the electric conductivity were done for Einstein-dilaton-three-Maxwell holographic model \cite{Arefeva:2020vae}. The dependence of the conductivity on the temperature, the chemical potential, the external magnetic field, and the spatial anisotropy of the heavy-ions collision (HIC) is studied. The electric conductivity jumps near the first-order phase transition are observed. This effect is similar to the jumps of holographic entanglement that were studied previously.
}

\begin{document}
\maketitle

\newpage
\section{Introduction}

The thermal-photon production in heavy-ion collisions plays an essential role in studying the quark-gluon plasma (QGP). Photons can be considered as probes of QGP because they do not interact with hadronic plasma. Experimental study of the production of thermal photons provides knowledge about many characteristics of QGP \cite{David:2019wpt}. In  particular,  the rate of photon production is related to the electric conductivity of QGP \cite{CaronHuot:2006te}. \\

To compare HIC experimental data to theoretical calculations, one needs non-perturbative calculations in QCD. 
A holographic approach is an effective tool for non-perturbative studies of QGP \cite{Solana, IA, DeWolf}. The present investigation is significant in the scope of recent experiments in high energy physics: FAIR and NICA projects. \\

There is rich literature devoted to holographic calculations of electric conductivity  
\cite{
Wu:2013qja,CaronHuot:2006te,Patino:2012py,Finazzo:2013efa,Arciniega:2013dqa,Iatrakis:2016ugz,Arefeva:2016rob,Avila:2021rcu}. The holographic calculations of electric conductivity  are related to the retarded correlator of currents in momentum space \cite{CaronHuot:2006te,Liu,Son:2007vk,
Son:2002sd,Policastro}. There are two different ways to find electric conductivity. In the first scheme \cite{Son:2007vk,Son:2002sd,Policastro,Iatrakis:2016ugz}, one calculates the retarded Green function using equations of motion. In the alternative approach, one uses the membrane paradigm \cite{Wilczek}. All previous calculations consider isotropic \cite{Iatrakis:2016ugz} or only partially anisotropic holographic models \cite{Patino:2012py,Wu:2013qja,Arefeva:2016rob,Avila:2021rcu},  in the later case only part  of coordinates enter the metric with different scale functions.
\\

The choice of different warp factors distinguishes isotropic and anisotropic holographic QCD models
\cite{Solana,IA,DeWolf,Kiritsis08,1301.0385,Donos:2014cya, yang2015,Dudal:2017max,Dudal:2018ztm,Gursoy:2018ydr, Mahapatra:2019uql,Bohra:2019ebj,Gursoy:2020kjd,He:2020fdi,Arefeva:2020vae, Zhou:2020ssi,Arefeva:2020byn,Ballon-Bayona:2020xtf,Bohra:2020qom,Rodrigues:2020ndy,Dudal:2021jav,Zhou:2021nbp,Caldeira:2021izy, Ballon-Bayona:2021tzw}. 
Many models establish the essential influence of anisotropy on the properties of QGP observables
 \cite{Giataganas:2012, AG,Ageev:2016gtl,AR-2018,Arefeva:2018cli,ARS-2019qfthep}. Note that there  are  many  studies  on  electric  conductivity  in inhomogeneous  condensed matter systems \cite{Ge:2015fmu,Khimphun:2016ikw,Kuang:2017rpx}. Also, account for non-zero chemical potential dramatically changes the phase structure of QCD \cite{Arefeva:2016rob,1301.0385,yang2015,Dudal:2017max,
Dudal:2018ztm, Mahapatra:2019uql, Arefeva:2020byn,Arefeva:2020vae,Gursoy:2020kjd,He:2020fdi,Rodrigues:2020ndy,Zhou:2021nbp,Zhou:2020ssi,Bohra:2019ebj,Bohra:2020qom,Caldeira:2021izy}. Previous studies have revealed close relation of anisotropic models with magnetic field and influence of magnetic field on observables in QGP are considered in  \cite{Wu:2013qja,Arciniega:2013dqa,Bohra:2019ebj,Gursoy:2018ydr,Rodrigues:2020ndy,Ballon-Bayona:2020xtf} and \cite{Arefeva:2020vae,He:2020fdi,Zhou:2020ssi,Avila:2021rcu, Dudal:2021jav}. \\

This paper studies direct photons moving through twice anisotropic QGP and investigates influence of chemical potential and magnetic field on the results. So, we aim at calculating the Green function and determining the electric conductivity tensor in twice anisotropic holographic theory\footnote{We call our model twice anisotropic since its holographic metric contains  two anizotropic parameters, see \eqref{metric-full-anis} and   \eqref{anisfun}.}. Following the two aforementioned prescriptions, we calculate the retarded Green's functions and demonstrate the agreement of these two approaches.
We chose the five-dimensional twice anisotropic holographic model for heavy quarks based on  Einstein-dilaton-three-Maxwell action  \cite{Arefeva:2020vae} for our calculations. The warp-factor in this model is chosen to reproduce the phase transitions structure for heavy quarks of the Columbia plot \cite{Brown:1990ev,Philipsen:2016hkv}. The first Maxwell field in action is related to chemical potential; the second describes spatial anisotropy, and the third accounts for an external magnetic field. In this model, a significant influence of the external magnetic field on the black hole solution and the confinement/deconfinement phase diagram was studied in \cite{Arefeva:2020vae}. In this model, we consider the impact of anisotropy, chemical potential and magnetic fields on electric conductivity.  \\

 We compare our results with lattice calculations. 
The electric conductivity was studied in isotropic lattice and in the vicinity of zero chemical potential \cite{Aarts:2014nba,Ghiglieri:2016tvj,Buividovich:2020dks,
Aarts:2020dda}.  Therefore, we compare our results with lattice calculations only for the isotropic ($\nu=1$) case with zero chemical potential ($\mu=0$). We reproduce the lattice data by varying the kinetic potential for the Maxwell field (see Fig.\ref{fig:Fit} below). With this universal kinetic function we calculate the DC conductivity for all other cases, i.e. anisotropic  collisions ($\nu>1$), non-zero chemical potential ($\mu>0$) and non-zero magnetic field. 
\\
  
The paper is organized as follows. In Sect.\ref{sec:Setup}, the holographic model and perturbation action are described. In Sect.\ref{Sect:conduct},  formulas for conductivity are obtained in twice anisotropic theory. 
 In Sect.\ref{Sect:numerical} numerical results are presented. In Sect.\ref{ConclDiscuss} we summarize the obtained results and link them to such quantities as the butterfly velocity, drag forces and the tension of the Wilson loop. 
Appendix \ref{Sect:green} shows the alternative calculations of QGP DC conductivity. The cumulative tables Table \ref{Table:1} and Table \ref{Table:2} in Appendix \ref{Sect:Tables} present plots for different set of parameters.

\section{Setup}\label{sec:Setup}

\subsection{Holographic model}

It was recognized that \cite{1105.3472, 1106.1637, 
JW,RS,AG,
  1811.11724} it is important to add anisotropy in the holographic
theory as QGP is an anisotropic media just after the HIC, and an
estimation for isotropisation time is about $1$--$5$ fm/c $\sim
10^{-24}$ s \cite{Strickland:2013uga}. One of motivations to deal with
anisotropic models is related with the problem of getting experimental
data for the energy dependence of the total multiplicity of particles
created in heavy-ion collisions \cite{Alice}. Isotropic holographic
models had not been able to reproduce the experimental multiplicity
dependence on energy (\cite{AG} and refs therein),
and to reproduce them anisotropic models of  Lifshitz type  with
a parameter $\nu$ were considered. It's value of about $\nu = 4.5$
gives the dependence of the produced entropy on energy in accordance
with the experimental data for the energy dependence of the total
multiplicity of particles created in heavy-ion collisions
(results by ATLAS and ALICE) \cite{ATLAS:2011ag,Alice}. Anisotropy is also related with
strong  magnetic fields typical for HIC \cite{0907.1396, 1103.4239, 1111.1949,1201.5108,Fotakis:2021diq}. 

We consider an anisotropic holographic model \cite{Arefeva:2020vae} 
 based on the action\footnote{This model is a generalization of \cite{AR-2018,AGG}}:

\bea  \label{action} 
    S_{bck} = \cfrac{1}{16\pi G_5} \int d^5x \ \sqrt{-g} &\cdot&
    \Big[ R - \cfrac{f^{(1)}(\phi)}{4} \ F^{(1)^2} 
      - \cfrac{f^{(2)}(\phi)}{4} \ F^{(2)}\,^2\\
      &\,\,\,&- \cfrac{f^{(B)}(\phi)}{4} \ F^{(B)}\,^2
      - \cfrac{1}{2} \ \partial_{\mu} \phi \partial^{\mu} \phi
      - V(\phi) \Big].
\nn 
  \eea 
  
 This action contains three different Maxwell fields with their own dilaton-dependent coupling functions\footnote{See \cite{Arefeva:2020vae} for discussion of the choice of coupling functions}  and each of these fields has its own interpretation. With  help of $F^{(1)}$ and $F^{(2)}$  the chemical potential and anisotropy are introduced, and $F^{(B)}$ describes the external magnetic field. 

 The ansatz for the metric is \be\label{metric-full-anis}
    ds^2=\frac{L^2 \mathfrak{b}(z)}{z^2} \left[-g(z)dt^2 +\mathfrak{g}_1(z)dx_1^2 +\mathfrak{g}_2(z)dx_2^2+\mathfrak{g}_3(z)dx_3^2 +\frac{dz^2}{g(z)} \right] , \\ 
\ee
where $L$ is the AdS-radius, $\mathfrak{b}(z)$ is the warp-factor, $g(z)$ is the blackening function, and we will consider $ \mathfrak{g}_i$ of the special forms
\begin{equation}
     \mathfrak{g}_1 = 1, \quad \mathfrak{g}_2 (z) = \left(\frac{z}{L}\right)^{2-\frac{2}{\nu }}, \quad \mathfrak{g}_3 (z) = e^ {  c_B z^2  } \left(\frac{z}{L}\right)^{2-\frac{2}{\nu }}. \label{anisfun}
\end{equation} 
The special form of prefactors
$\left(\frac{z}{L}\right)^{2-\frac{2}{\nu }}$  describes non-symmetry of heavy-ion collision (HIC), and $c_B$ is the coefficient of secondary anisotropy related to the external magnetic field. The specific choice of
the warp-factor $\mathfrak{b}(z)$ determines the phase transitions structure of the model.   The warp-factor of the form $\mathfrak{b}(z)=e^{-cz^2/2}$ reproduces the heavy quarks' phase transitions structure of the Columbia plot \cite{Brown:1990ev,Philipsen:2016hkv,Aarts:2014nba,
Ghiglieri:2016tvj}. The warp-factor $\mathfrak{b}(z) = e^{- 2 a \ln (bz^2 + 1)}$ describes the light quarks' case. The explicit solutions for the blackening function $g$, the gauge kinetic functions $f_0$, $f_2$ and $f_B$, the dilaton field, and the dilaton field potential $V$ are given in \cite{Arefeva:2020vae}.

\begin{figure}[h]
\center{\includegraphics[width=0.7\linewidth]{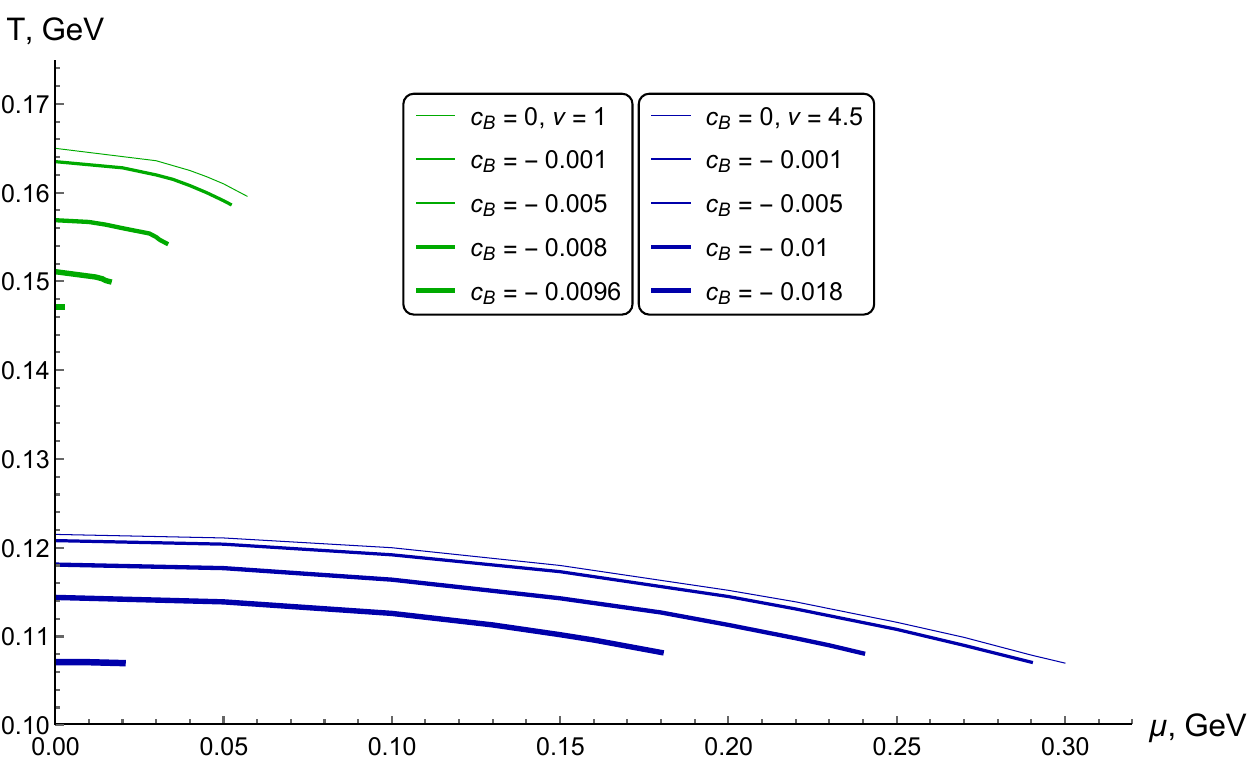}}
\caption{The BB phase transitions structure from \cite{Arefeva:2020vae} for different anisotropy parameter $\nu$ and magnetic field's parameter $c_B$.}
\label{Fig:flow}
\end{figure}
This model's phase transition structure is presented in 
Fig.\ref{Fig:flow}. A remarkable feature of Einstein-dilaton-Maxwell models is the Van der Waals type of the temperature dependence on the size of horizon $z_h$. The temperature function can be three-valued for some values of parameters (chemical potential, anisotropy, magnetic field etc.). The Hawking-Page phase transition between AdS Black Hole and thermal AdS occurs in these models for zero chemical potential. For any non-zero chemical potential, there is a phase transition between small and large black holes. Even for zero chemical potential but non-zero magnetic field, the phase transition becomes the phase transition between small and large black holes (BB). For these phase transitions, such values as entropy, holographic entanglement entropy, and baryon density can experience a jump \cite{Arefeva:2020uec,Arefeva:2020byn}. In this work we observe the jump of electric conductivity at the temperature of BB phase transition, see next Sect.\ref{Sec:NumRes}.

To investigate properties of direct photons in heavy-ion collisions using holographic duality one has to introduce one more Maxwell field \cite{Wu:2013qja,CaronHuot:2006te,Patino:2012py,Finazzo:2013efa,Arciniega:2013dqa,Erdmenger,Iatrakis:2016ugz,Arefeva:2016rob,Avila:2021rcu,Liu,Wilczek}. It is described  by an action  
\begin{equation}\label{probe action}
     S_{out} = -\frac{1}{4}\int d^5x\sqrt{-g}f_0F_{MN}F^{MN} ,
\end{equation}
here
 $f_0=f_0(\phi)$ is the function of coupling of the Maxwell field to the dilaton (also called the gauge kinetic function) and we use capital Latin letters to number components of 5-dim objects $M=0,1,2,3,4$ and Greek letters for objects on 4-dim Minkowski spacetime $\mu=0,1,2,3$.  The choice of $f_0$ function allows one to fit lattice results, see below Sect.\ref{Sect:f_0 neq 1}. 
Proceeding with a variational procedure for $S_{out}$, one finds
\begin{align}
    \delta S_{out} = -\int_{\partial \Omega}d\omega \sqrt{-g}f_0n_MF^{MN}\delta A_N +\int_{\Omega} d^5x \partial _M(\sqrt{-g}f_0F^{MN})\delta A_N=0,
\end{align}
where the manifold $\partial \Omega$ we integrate over is the 4-dim Minkowski space-time, $d\omega=d^4x$ is measure on the aforementioned manifold and $n_M$ is the outward unit normal vector to the boundary: $n_M=(0,0,0,0,1)$ on the horizon $z=z_h$ and $n_M=(0,0,0,0,-1)$ on the boundary $z=0$. Therefore, to get rid of surface terms one adds to $S_{out}$ the surface term $S_{sur\!f}$ and considers it as an addition to the initial action of the theory. The surface term in this case is  
\begin{equation}
    S_{sur\!f}=\left. \int d^4x \sqrt{-g}f_0F^{4\mu} A_{\mu} \right| ^{z=z_h}_{z=0}.\label{Surface}
\end{equation}
This form of $S_{sur\!f}$ is essential in  the so-called 'membrane paradigm' \cite{Wilczek,Liu}. Then equations of motion are just Maxwell's equations
\begin{equation}
    \partial _M(\sqrt{-g}f_0F^{MN})=0.
\end{equation}

\subsection{Direct photons emission rate and electric conductivity}
The number of photons emitted per unit time per unit volume $\Gamma$ (photon emission rate) in thermal equilibrium is given by the light-like correlator \cite{Kapusta:2006pm, Erdmenger}
\bea
d \Gamma=-\frac{d \mathbf{k}}{(2 \pi)^{3}} \frac{e^{2} n_{b}(|\mathbf{k}|)}{|\mathbf{k}|} \operatorname{Im}\left[ \eta_{\mu \nu} G^{\mu \nu}_R  \right] _{k^{0}=| \mathbf{k} |} ,
\eea
where $$n_b(|\mathbf{k}|)=\frac{\mathcal{A}}{\e ^{-|\mathbf{k}|/T}-1}$$ is Bose-Einstein thermal distribution function, $\eta_{\mu \nu}$ is Minkowski metric tensor, photon's 4-momentum is $k^\mu=(k^0,\mathbf{k})$. 
Note that the retarded Green's function $G_R^{\mu \nu}$ is related to the electric conductivity through the Kubo relation $$\sigma^{\mu \nu}=-\frac{G_R^{\mu \nu}}{i w},$$ so the direct photons emission rate is connected to the conductivity of QGP.

\section{Electric conductivity in twice anisotropic background}
\label{Sect:conduct}
To find the electric conductivity, we add a probe Maxwell field \eqref{probe action} to \eqref{action}.
\label{Sect:ansatz}
Consider an ansatz for this probe field in the form of a plane wave propagating in $x_3$ direction
\be
    A_M({t},{x_3},{z}){=}\psi _M(z) \exp (-i (wt-k {x_3})),\qquad
    M=0,...4.
\ee
From the equation of motion for $A_4(t, x_3, z)$ (from this moment we write $A_M = A_M({t},{x_3},{z})$ for shorthand) component it follows that
\begin{equation}
    \psi _4(z)\text{=}i\frac{-k g(z) \psi _3'(z)-w \mathfrak{g}_3(z) \psi _0'(z)}{k^2 g(z)-w^2 \mathfrak{g}_3(z)},
\end{equation}
where $'$ denotes derivative with respect to $z$.

We also introduce notation for the longitudinal component of electric field
\begin{equation}
     E_3=F_{03}=wA_3+kA_0
\end{equation}
and for two transverse components as
\begin{equation}
    E_i=wA_i, \qquad i=1,2.
\end{equation}
For brevity we drop out arguments of the blackening function $g$, anisotropy functions $\mathfrak{g}_1,\, \mathfrak{g}_2,\, \mathfrak{g}_3$, the gauge kinetic function $f_0$ and the warp-factor $\mathfrak{b}$ in the following equations. 
Thus equations of motions are 

\bea\label{eq: Ee}
    &&E_3'' + E_3' \left(\dfrac{\fb'}{2 \fb}+\dfrac{f_0'}{f_0}-\dfrac{w^2 \mathfrak{g}_3 g'}{k^2 g^2-w^2 g \mathfrak{g}_3}+\dfrac{w^2 \mathfrak{g}_3'}{k^2 g-w^2 \mathfrak{g}_3}+\dfrac{\mathfrak{g}_1'}{2 \mathfrak{g}_1}+\dfrac{\mathfrak{g}_2'}{2 \mathfrak{g}_2}+\dfrac{\mathfrak{g}_3'}{2 \mathfrak{g}_3}-\dfrac{1}{z}\right) +\label{eq: Eel} \nonumber\\
    &&+E_3\dfrac{w^2 \mathfrak{g}_3-k^2 g}{g^2 \mathfrak{g}_3}=0 ;\\
    &&E _1'' + E _1' \left(\dfrac{\fb'}{2 \fb}+\dfrac{f_0'}{f_0}+\dfrac{g'}{g}+\dfrac{\mathfrak{g}_2'}{2 \mathfrak{g}_2}+\dfrac{\mathfrak{g}_3'}{2 \mathfrak{g}_3}-\dfrac{\mathfrak{g}_1'}{2 \mathfrak{g}_1}-\dfrac{1}{z}\right)+\label{eq: Ee1} E _1\dfrac{w^2 \mathfrak{g}_3-k^2 g}{g^2 \mathfrak{g}_3}=0 ;\\
    &&E _2'' + E _2' \left(\dfrac{\fb'}{2 \fb}+\dfrac{f_0'}{f_0}+\dfrac{g'}{g}+\dfrac{\mathfrak{g}_1'}{2 \mathfrak{g}_1}+\dfrac{\mathfrak{g}_3'}{2 \mathfrak{g}_3}-\dfrac{\mathfrak{g}_2'}{2 \mathfrak{g}_2}-\dfrac{1}{z}\right)+\label{eq: Ee2} E _2 \dfrac{w^2 \mathfrak{g}_3-k^2 g}{g^2 \mathfrak{g}_3}=0 .
\eea

Then the on-shell action is 
\be
    S_{sur\!f}=\int \frac{d^4 k}{(2\pi)^4} \frac{f_0 \,g}{z} \left(  \left. \mathcal{E}^*_1\sqrt{\frac{\fb\, \fg_3 \fg_2}{\fg_1}} \mathcal{E}'_1+\mathcal{E}^*_2\sqrt{\frac{\fb \, \fg_3 \fg_1}{\fg_2}}\mathcal{E}'_2-\mathcal{E}_3^*\frac{\sqrt{\fb \, \fg_1 \fg_2 \fg_3}}{\frac{k^2}{w^2}g-\fg_3}\mathcal{E}'_3  \right) \right| _{z=0}^{z=z_h} ,
\ee
where $\mathcal{E}_i={E_i}/{w}$ for $i=1,2,3$.

Following \cite{Liu, Iatrakis:2016ugz} we make the on-shell action quadratic by introducing a new variable $\zeta_3$ proportional to canonical momentum of field $\mathcal{E}_3$
\begin{equation}
    \zeta_3=-\frac{\mathcal{E}'_3}{\mathcal{E}_3}\frac{2f_0 g}{z}\frac{\sqrt{\fb \, \mathfrak{g}_1 \mathfrak{g}_2 \mathfrak{g}_3 }}{\mathfrak{g}_3-g\frac{k^2}{w^2}}.
\end{equation}
For further convenience we define
\begin{equation}
    B_3 \equiv \frac{2f_0 g}{z}\frac{\sqrt{\fb \, \mathfrak{g}_1 \mathfrak{g}_2 \mathfrak{g}_3 }}{\mathfrak{g}_3-g\frac{k^2}{w^2}}.
\end{equation}
For the longitudinal direction the equation of motion \eqref{eq: Eel} becomes
\begin{equation}
    \frac{\mathcal{E}_3''}{\mathcal{E}_3} = \zeta_3 \frac{B_3'}{B_3^2}-\frac{w^2 \mathfrak{g}_3-k^2 g}{g^2 \mathfrak{g}_3}.
\end{equation}
Taking derivative of $\zeta_3$ with respect to $z$
\begin{equation}
    \zeta_3'=-\frac{\mathcal{E}_3''}{\mathcal{E}_3}B_3+\frac{B_3'}{B_3}\zeta_3+\frac{\zeta_3^2}{B_3}=-\frac{\zeta_3 B_3'}{B_3}+B_3\frac{w^2 \mathfrak{g}_3-k^2 g}{g^2 \mathfrak{g}_3}+\frac{\zeta_3 B_3'}{B_3}+\frac{\zeta_3^2}{B_3}.
\end{equation}
Finally, the differential equation on $\zeta_3$ is 
\begin{equation}
    \left( \frac{\zeta_3}{w} \right) ' -\frac{f_0\sqrt{\fb \, \mathfrak{g}_1 \mathfrak{g}_2 \mathfrak{g}_3} }{\mathfrak{g}_3-g\frac{k^2}{w^2}}\frac{w}{z g}\left[ \left( \frac{\zeta_3}{w} \right) ^2 \left( \frac{\mathfrak{g}_3-g\frac{k^2}{w^2}}{2f_0 \sqrt{\fb \, \mathfrak{g}_1 \mathfrak{g}_2 \mathfrak{g}_3}}z \right) ^2+\frac{\mathfrak{g}_3-g\frac{k^2}{w^2}}{\mathfrak{g}_3} \right] =0 \label{e3}
\end{equation}
We require $\zeta_3'(z)\neq \infty$ for all $z$ since this quantity is related to observable values. The blackening function vanishes on the horizon, making the second term infinite. We are performing series expansion for both the blackening function \eqref{gonhor} and the function inside square brackets (denote it $U(z)$ for convenience) near the horizon
\bea
    U(z) &=& \left( \dfrac{\zeta_3}{w} \right) ^2 \left( \dfrac{\mathfrak{g}_3-g\frac{k^2}{w^2}}{2f_0 \sqrt{\fb\, \mathfrak{g}_1 \mathfrak{g}_2 \mathfrak{g}_3}}z \right) ^2+\dfrac{\mathfrak{g}_3-g\frac{k^2}{w^2}}{\mathfrak{g}_3}= \nn \\
    &=& U(z_h)+U'(z_h)(z-z_h)+o(z-z_h)^2.
\eea
Clearly, the pole at $z=z_h$ cancels if $U(z_h)=0$. This requirement translates into boundary condition for the Cauchy problem \eqref{e3}
\begin{equation}
    \zeta_3(z_h)= 2iw\frac{f_0(z_h)}{z_h}\sqrt{\frac{\fb(z_h) \mathfrak{g}_1(z_h) \mathfrak{g}_2(z_h) }{\mathfrak{g}_3(z_h)}}.
\end{equation}
In the low-frequency limit $w\rightarrow0$, the equation \eqref{e3} simply becomes 
\begin{equation}
    \zeta_3'(z)=0.
\end{equation}
Therefore, the solution for all values of $z$ is 
\begin{equation}
    \zeta_3=2iw\frac{f_0(z_h)}{z_h}\sqrt{\frac{\fb(z_h) \mathfrak{g}_1(z_h) \mathfrak{g}_2(z_h) }{\mathfrak{g}_3(z_h)}}.
\end{equation}
Using the Kubo formula $\sigma^{\mu \nu} = -G^{\mu \nu}_R/iw$ we obtain the  33-component of QGP DC  conductivity tensor (as usually, the low-frequency conductivity is called  DC  conductivity)
\be\label{sigma33}
    \sigma^{33}=\frac{2 f_0(z_h)}{z_h}\sqrt{\frac{\fb(z_h) \mathfrak{g}_1(z_h) \mathfrak{g}_2(z_h) }{\mathfrak{g}_3(z_h)}}.
\ee
Doing all the same we obtain the 11 and 22 components of QGP DC  conductivity
\begin{equation}\label{sigma11}
  \sigma^{11} = \frac{2f_0(z_h)}{z_h}\sqrt{\frac{\fb(z_h)\mathfrak{g}_3(z_h) \mathfrak{g}_2(z_h)}{\mathfrak{g}_1(z_h)}},
\end{equation}
\begin{equation}\label{sigma22}
   \sigma^{22} = \frac{2f_0(z_h)}{z_h}\sqrt{\frac{\fb(z_h)\mathfrak{g}_3(z_h) \mathfrak{g}_1(z_h)}{\mathfrak{g}_2(z_h)}}.
\end{equation}
Note, that these results also agree with isotropic \cite{Iatrakis:2016ugz} and partially anisotropic \cite{Arefeva:2016rob} cases. The same results can be obtained using another prescription proposed by Son and Starinets in \cite{Son:2002sd}. One can find detailed calculations in Appendix \ref{Sect:green}. 


\section{Numerical Results}\label{Sec:NumRes}
\label{Sect:numerical}
Formulas \eqref{sigma33}, \eqref{sigma11} and \eqref{sigma22}  show, that components of conductivity are defined by functions $\fb$, $f_0$ and  $\mathfrak{g}_i$, $i=1,2,3$ at the horizon.  

\subsection{Conductivity asymptotic behaviour. }
\label{Sect:f_0 = 1}
We take $f_0(z)=1$ in this subsection. Following \cite{Arefeva:2020vae}, we make calculations in our model at fixed $L=1$ and $c=0.227$, and taking the warp-factor in the string frame
\begin{equation}
    \fb_{s}(z) = \exp \left(-\frac{1}{2} c z^2+\sqrt{\frac{2}{3}} \phi\right) .
\end{equation}

\begin{figure}[h!]
\center{\includegraphics[width=0.4\linewidth]{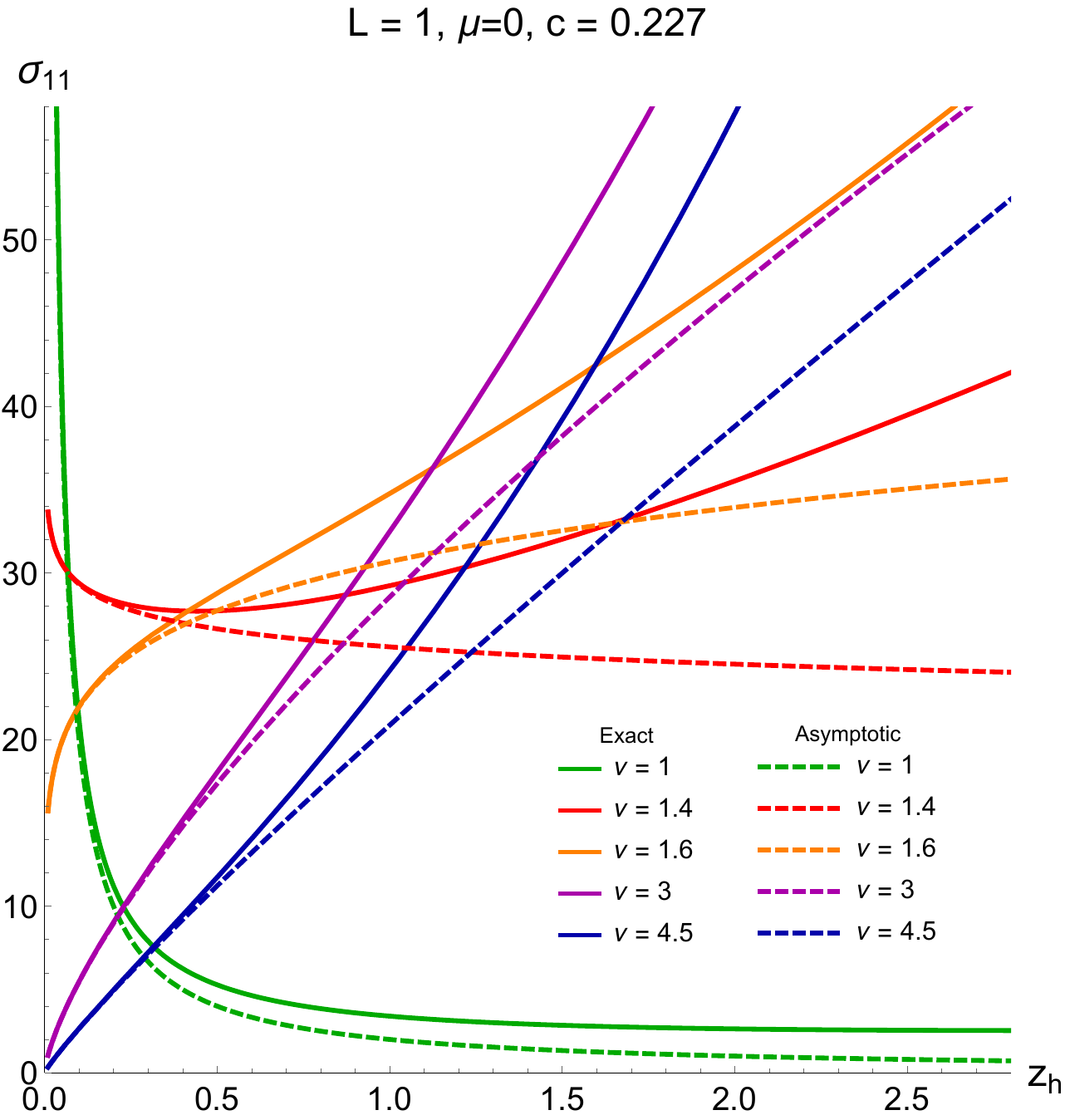}\qquad\qquad\includegraphics[width=0.4\linewidth]{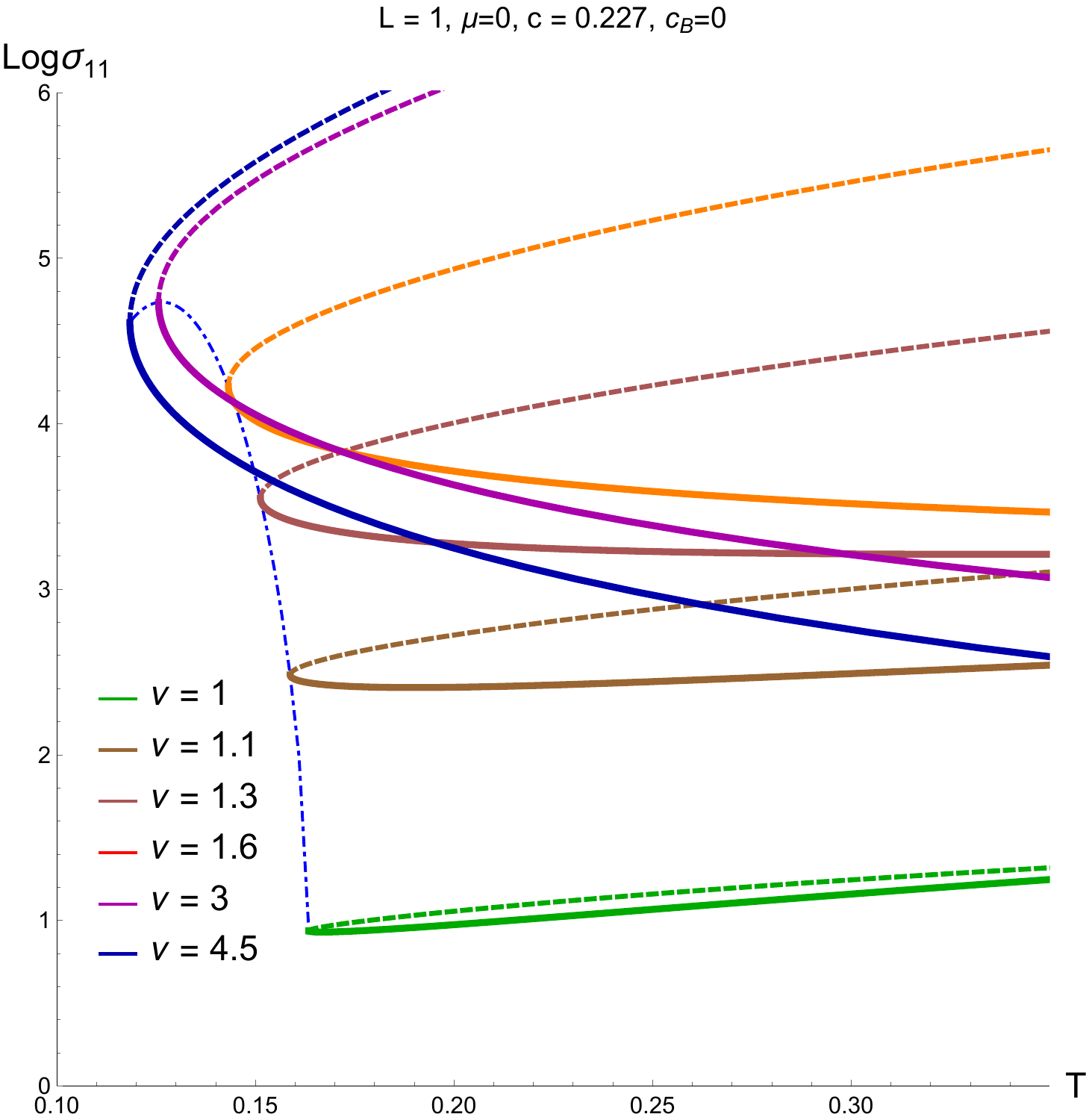}\\A\qquad \qquad\qquad\qquad\qquad\qquad \qquad B}
\caption{A) The dependence of the electric conductivity $\sigma^{11}$ on the size of horizon $z_h$ for different values of anisotropy parameter $\nu$ and zero magnetic field $c_B=0$. The solid lines show exact calculations, while dashed correspond to asymptotics around $z_h=0$. B) The dependence of electric conductivity $\sigma^{11}$ on the temperature $T$ for different values of anisotropy parameter $\nu$ and $\mu=0$ in logarithmic coordinates. The blue dot-dashed line presents the Hawking-Page phase transition points. Dashed lines represent values of $\sigma^{11}$ calculated in thermodynamically unstable phase.}
\label{fig3}
\end{figure}

In Fig.\ref{fig3}.A the electric conductivity $\sigma^{11}$ as a function of the size of the horizon $z_h$ is presented for different values of anisotropy parameter $\nu$ and zero magnetic field $c_B=0$. The asymptotic behaviour near $z_h=0$ is plotted in dashed curves and given by the following expression
\begin{align}
    \sigma^{11} \sim z_h^{\frac{6 \nu +\sqrt{6} \sqrt{\nu -1}-6}{3 \nu }-1}.\label{sigma_11_asymp}
\end{align}
One can see that around zero $z_h$, this power law exhibits two different regimes for $\nu$ below and above $\approx 1.457$. For small anisotropy ($\nu<1.457$), we have diverging at the origin and monotonically decreasing $\sigma^{11}$. For large anisotropy ($\nu>1.457$), the conductivity becomes monotonously increasing. Later this feature will give rise to peculiar thermodynamic properties of DC conductivity.

In Fig.\ref{fig3}.B the dependence of electric conductivity $\sigma^{11}$ on the temperature $T$ for $\mu=0$ and different values of anisotropy parameter $\nu$ is shown in logarithmic coordinates. In this case the electric conductivity $\sigma^{11}$ exists only in a certain temperature range due to phase transitions structure \cite{Arefeva:2020vae}. The Hawking-Page transition points are the leftmost points of curves when chemical potential is zero. For temperatures below the Hawking-Page transition, the conductivity should be calculated in thermal AdS. $\sigma^{11}$ on this plot is presented for calculations in the AdS black hole. We expect a phase transition at $T=T_{HP}$. The blue dot-dashed line presents the Hawking-Page phase transition points. Dashed lines represent values of $\sigma^{11}$ calculated in thermodynamically unstable phase.

Unlike $\sigma^{11}$, the 33-component of conductivity does not exhibit the change of asymptotic behavior near the horizon. Instead of \eqref{sigma_11_asymp}, one has near $z_h=0$
\be
    \sigma^{33}\sim z_h^{\frac{0.816\sqrt{\nu-1}}{\nu}-1}. \label{sigma_33_asymp}
\ee 
The power of $z_h$ is always negative for $\nu \in [ 1,4.5 ]$ taking the maximum value of $-0.592$ for $\nu=2$. Therefore, $\sigma^{33}(z_h)$ monotonously decreases for all considered values of $\nu$.  

Plots of $\sigma^{11}(T)$ and $\sigma^{33}(T)$ are presented in Table \ref{Table:1} for different values of magnetic field parameter, chemical potential and anisotropy. We would like to mention the discontinuous behaviour of the conductivity as a function of temperature. We observe a jump at the temperature of Hawking-Page phase transition, which disappears as the magnetic field and chemical potential increase.


\subsection{The  lattice data fit for zero chemical potential}\label{Sect:f_0 neq 1}

\begin{figure}[b!]
\center{\includegraphics[width=0.5\linewidth]{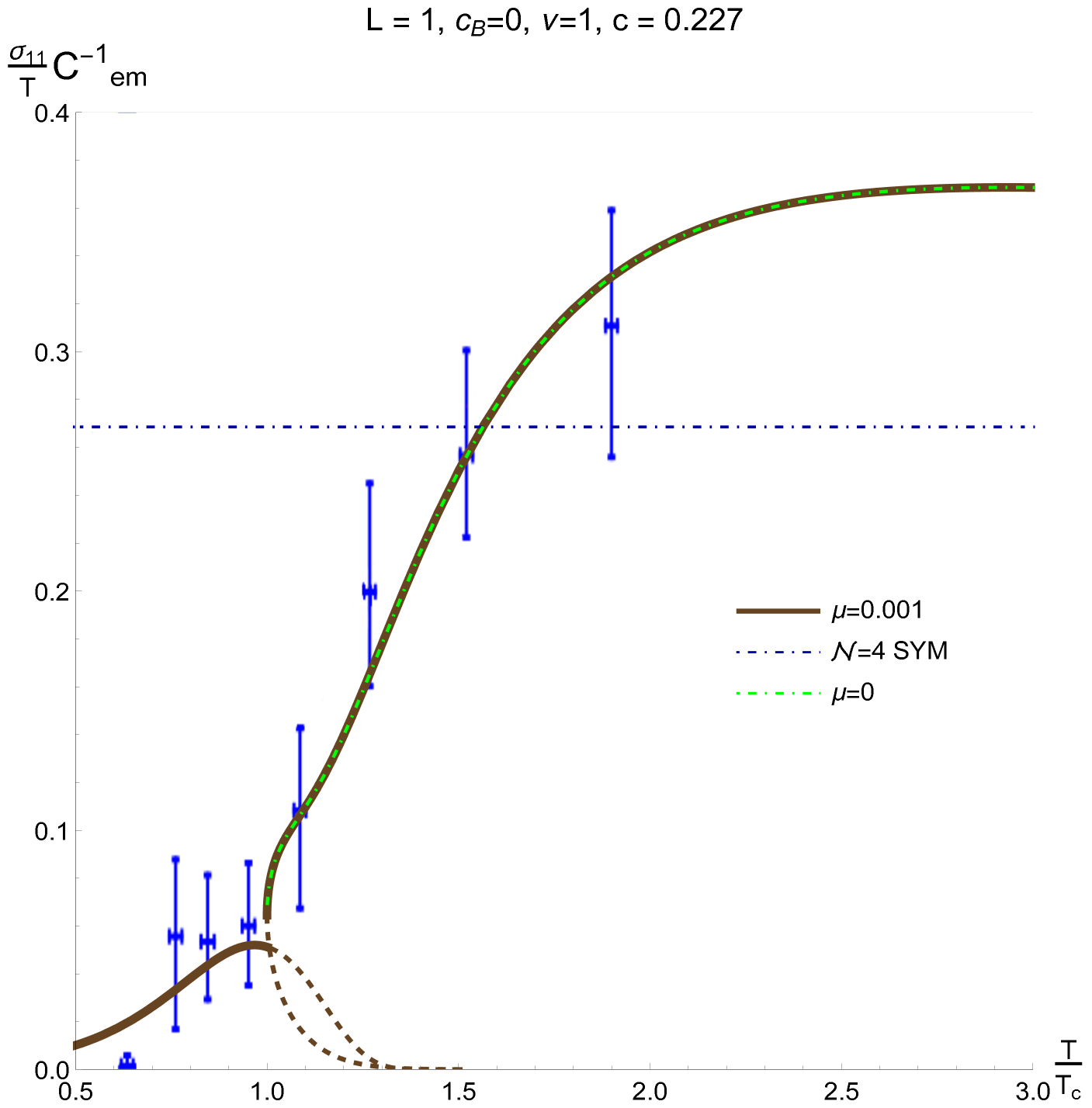}}
\caption{The dependence of ratio $\sigma^{11}/T$ on the temperature $T$ normalized by the critical value $T_c=0.163$ corresponding to the Hawking-Page phase transition is presented. The solid brown line corresponds to the stable phase with a small non-zero chemical potential $\mu=0.001$ while the green dash-dotted line represents the $\mu=0$ case.
The blue dash-dotted line is $\mathcal{N}=4$ SYM conductivity $\sigma/T=e^2N_c^2/ 16 \pi $.  Blue dots correspond to lattice calculations from \cite{Aarts:2014nba} for $N_c=N_f=3$. }
\label{fig:Fit}
\end{figure}

To fit the lattice results for zero chemical potential, we use the gauge kinetic function $f_0(\phi)$. This function couples Maxwell field to dilaton and depends on the size of horizon $z_h$, anisotropy parameter $\nu$ and magnetic field parameter $c_B$. The choice of this function is purely phenomenological. To get the fit presented in Fig.\ref{fig:Fit}, we take the kinetic function $f_0$ as
\bea
    f_0(\phi)&=& \frac{1}{440}  \left\{  \left( \exp \left[ -\frac{1}{10} \left( \phi (z_h,\nu , c_B)^3+\phi (z_h,\nu , c_B) \right) \right] + 0.3 \exp \left[ -\frac{1}{30}  \phi (z_h,\nu , c_B)^2 \right] \right) + \right. \nn\\
    &&\left.\qquad +\frac{1}{15}\exp \left[ -\frac{1}{0.04}  \phi (z_h-z_h^{s},\nu , c_B)^2 \right]  \right\} ,
    \label{fitfunc}
\eea
where $z_h^s$ is a value that shifts the last Gaussian term to the point, where stable and unstable phases meet each other. It is necessary because two branches of the stable phase are widely separated in terms of $z_h$ but close to each other as functions of $T$. 

In Fig.\ref{fig:Fit}  the dependence of  ratio $\sigma^{11}/T$ on the temperature $T$ normalized by the critical value $T_c=0.163$ corresponding to the Hawking-Page phase transition is presented. The solid brown lines correspond to the stable phase with a small non-zero chemical potential $\mu=0.001$ while the green dash-dotted line represents the $\mu=0$ case. One can see that these two curves coincide in the region $T>T_c$, but the green one is not defined below $T_c$. 
The blue dash-dotted line is $\mathcal{N}=4$ SYM conductivity $\sigma/T=e^2N_c^2/ 16 \pi $. $C_{em}^{-1}=\dfrac{2e^2}{3}$ is electromagnetic constant for number of flavors $N_f=3$ and colors $N_c=3$. Blue dots with error bars correspond to lattice calculations from \cite{Aarts:2014nba} for $N_c=N_f=3$. Although the lattice calculations are made for light quarks (u,d,s), the pion mass is $M_{\pi}=384$ MeV, which is almost 3 times bigger than its physical mass. In addition, the ratio $M_{\pi}/M_{\rho}$ is 3 times bigger. This obstacle indicates that there are yet no reliable lattice results for both heavy and light quarks. Either way, we use these data to adjust the coupling $f_0(\phi)$ so that in a simple case of vanishing chemical potential and zero magnetic field, the electric conductivity in our model fits lattice results. We have already mentioned that our model's ansatz corresponds to the case of heavy quarks, so in general, $\sigma$ does not have to agree with these data. Once $f_0(\phi)$ is tuned, it is possible to predict the behaviour of $\sigma$ for different parameters of the magnetic field, anisotropy and chemical potential. In this work, we define the dilaton coupling as in \eqref{fitfunc}.

\begin{figure}[h!]
\centering
\includegraphics[width=0.25\linewidth]{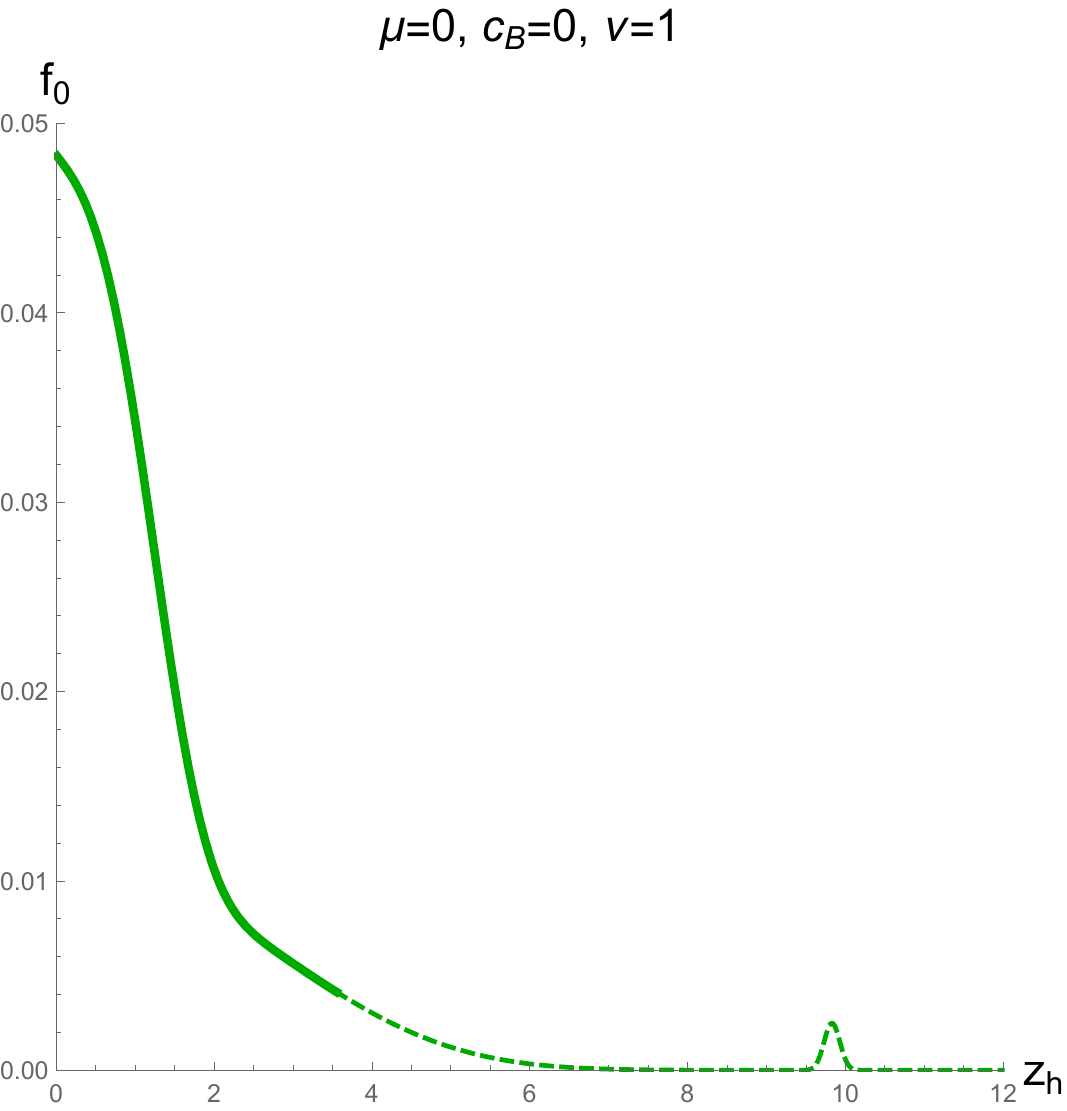}\qquad\qquad\qquad\includegraphics[width=0.25\linewidth]{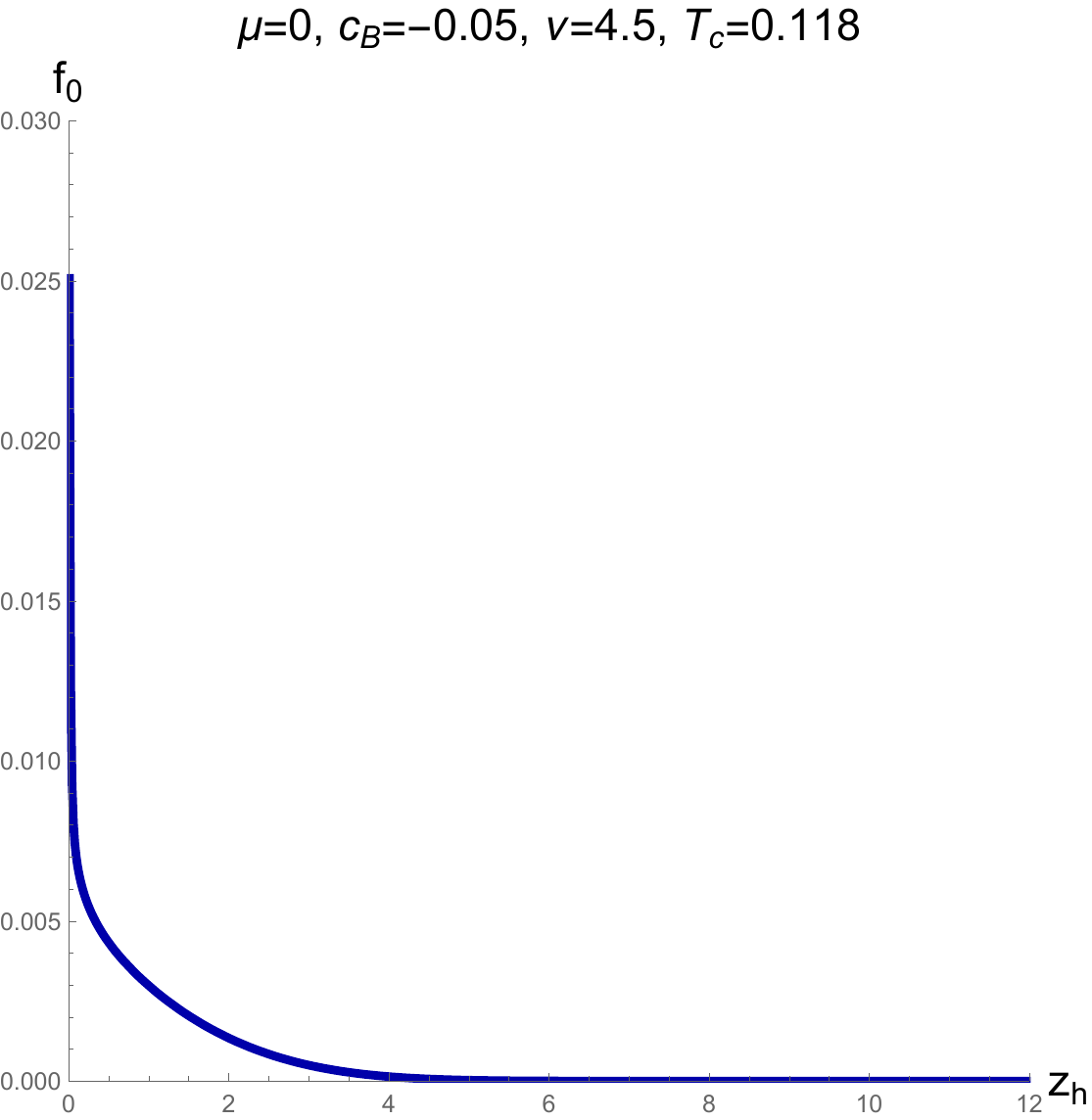} \\
A \hspace{70 mm} B\\\,\\
\includegraphics[width=0.25\linewidth]{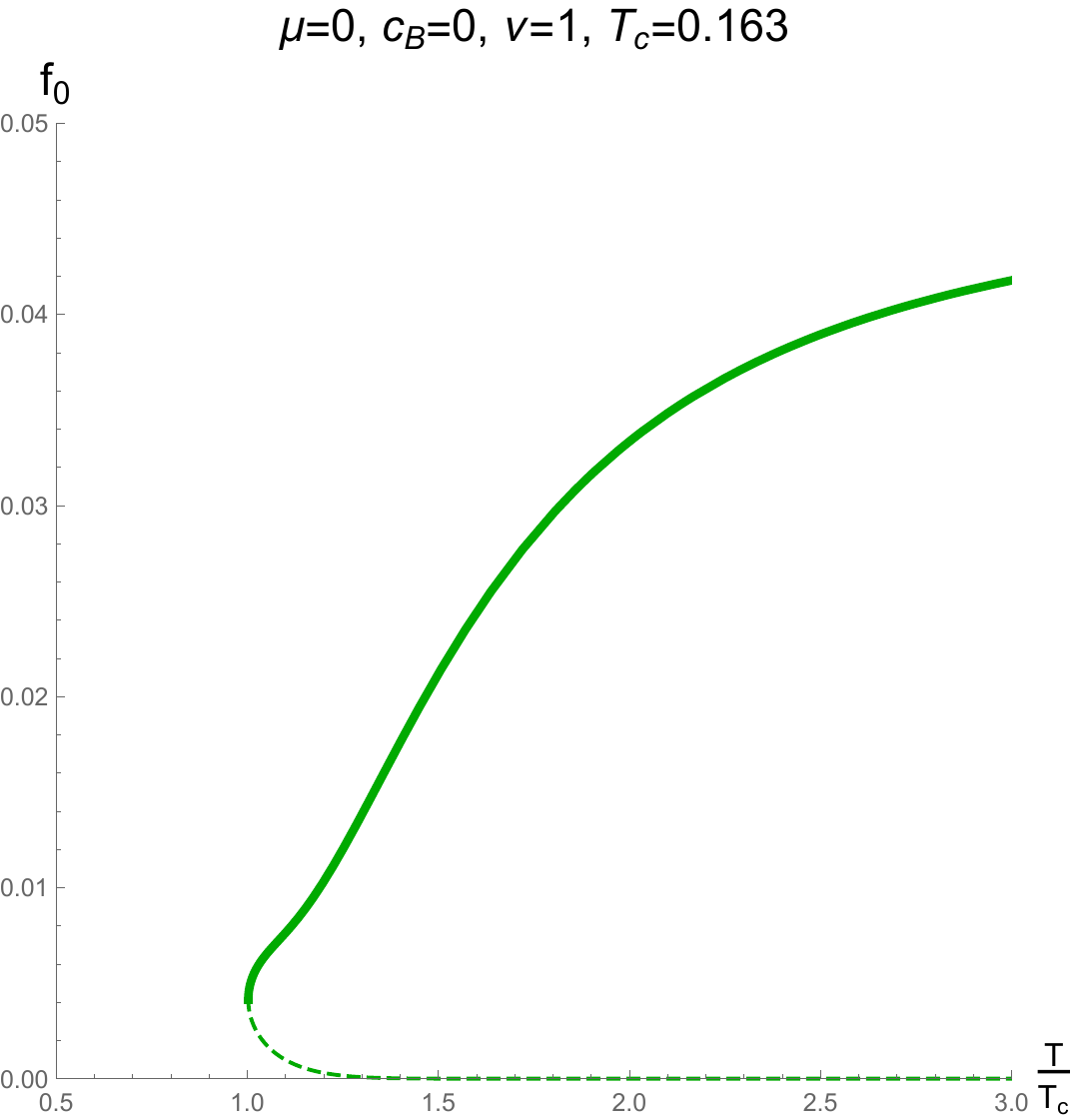}
\qquad\qquad\qquad\includegraphics[width=0.25\linewidth]{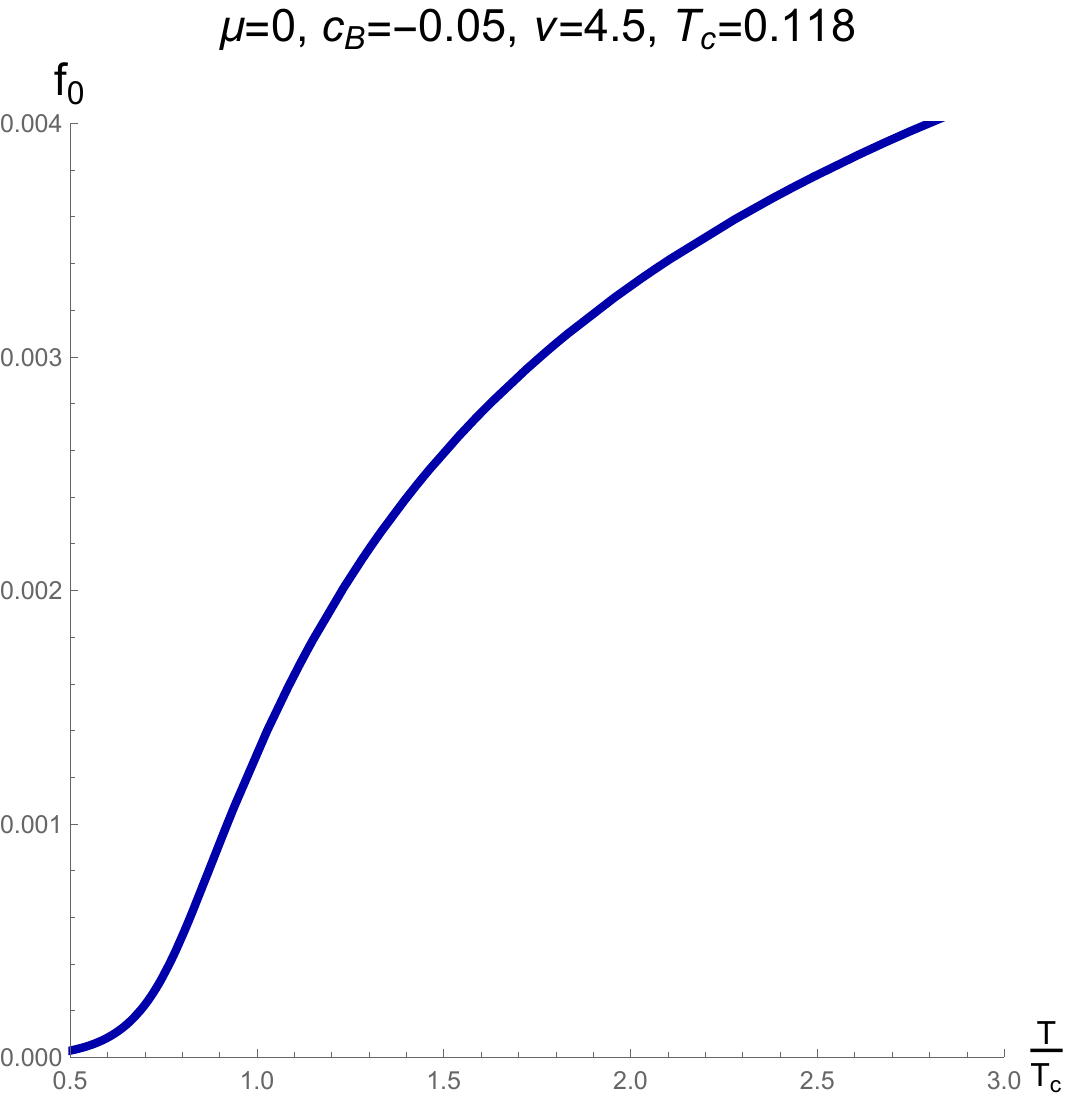}
\\
C \hspace{70 mm} D 
\caption{The dependence of our choice of $f_0$ function \eqref{fitfunc} on the size of the horizon  (A,B) and on the  temperature $T/T_c$ (C,D) for isotropic case (green curves) and anisotropic one (dark blue curves).}
\label{fig:f0func}
\end{figure}

Plots of $f_0$ as a function of both $z_h$ and $T$ are presented on Fig.\ref{fig:f0func} for different sets of parameters. As one can see on Fig.\ref{fig:f0func}.A there is a hump near $z_h\approx 10$. It corresponds to the rise of the stable phase of $\sigma^{11}/T$ in the $\frac{T}{T_c}\in [0.5 , 1.0]$ region in Fig.\ref{fig:Fit}. Also note, that $f_0$ does not depend on the chemical potential $\mu$. Thus, all the model's predictions about the impact of finite chemical potential are independent of a particular choice of kinetic function.

\subsection{Quadratic approximation of conductivity dependence for small  chemical potential}

\begin{figure}[h]
\center{\includegraphics[width=0.3\linewidth]{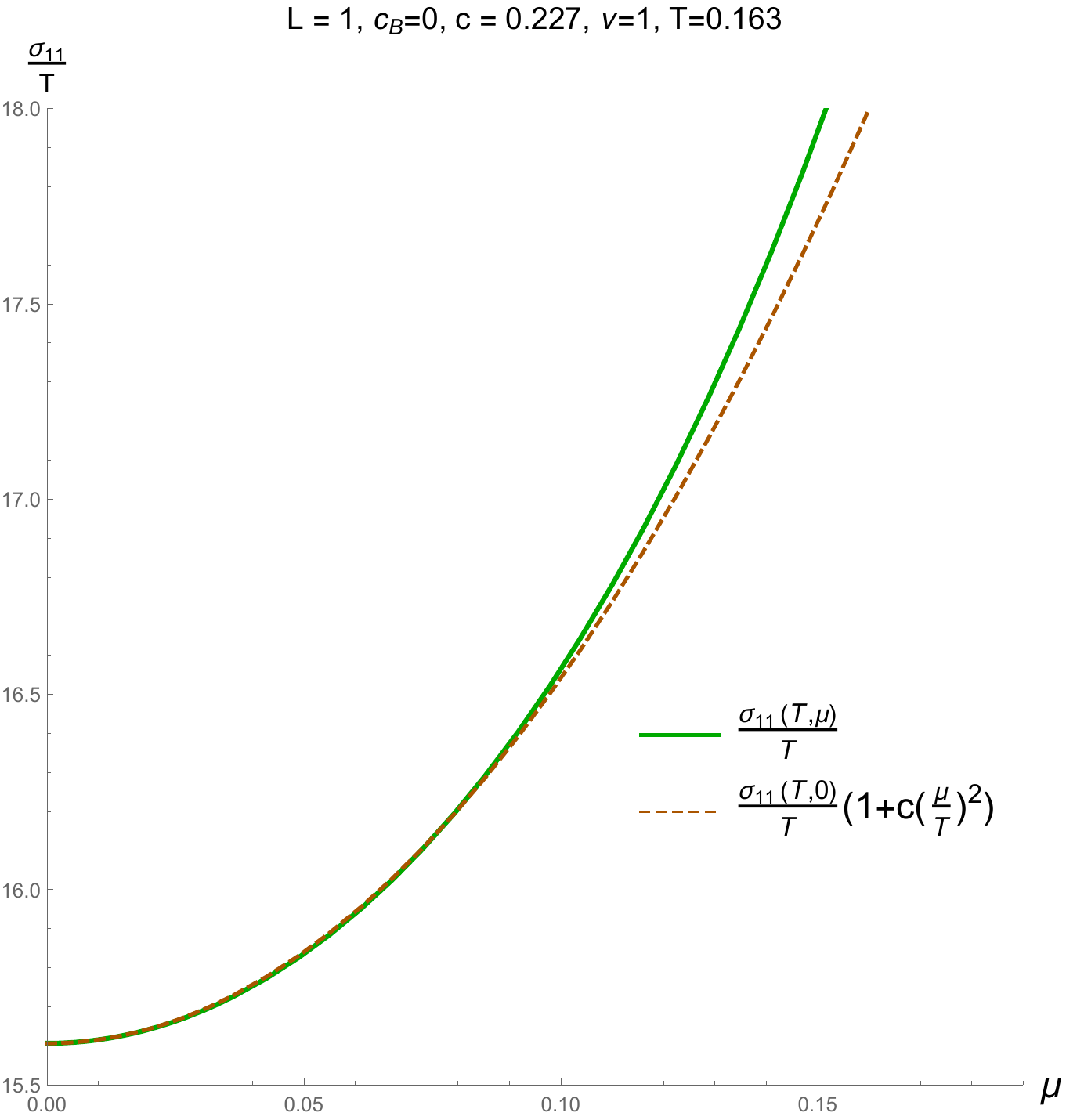}}
\caption{The dependence of $\sigma^{11}/T$ on the chemical potential $\mu$ for temperature $T=0.163$ corresponding to the Hawking-Page phase transition. The green curve represents the exact answer, the dashed brown curve is the power series in $\mu$ around $\mu=0$.}
\label{fig:Chem}
\end{figure}
Symmetries of the QCD action imply the evenness of $\sigma/T$ as a function of chemical potential. Therefore its power series may contain only even powers of $\mu$: \be\dfrac{\sigma(T, \mu)}{T}=\dfrac{\sigma(T, 0)}{T}\left(1+c(T)\left(\dfrac{\mu}{T}\right)^{2}+O\left(\mu^{4}\right)\right).
\ee
This quadratic dependence has been studied in numerous works using lattice calculations \cite{Buividovich:2020dks}, and different phenomenological models \cite{Fotakis:2021diq}. It was shown, that the function $c(T)$ remains constant for quite a wide range of temperatures around the phase transition. In our case, the best match of the two curves occurs for $c(T) \approx 0.16$, which agrees with \cite{Buividovich:2020dks}. The plot Fig.\ref{fig:Chem} shows an adequate behaviour of the electric conductivity in our model, consistent with the conventional understanding of this phenomenon.  This plot is also a  clear illustration of independence of the function $f_0$ of chemical potential. 
It follows from the Einstein equations for model \cite{Arefeva:2020vae} that the dilaton field does not depend on the chemical potential. Since the gauge kinetic function $f_0(\phi)$ depends on the dilaton only, the $f_0$ in the ratio $\sigma /T$ in Fig.\ref{fig:Chem} remains constant for all values of $\mu$. Thus, the qualitative behaviour is right despite of the particular choice of the gauge kinetic function.

\newpage
\subsection{Calculation of $\sigma^{11}$}


\begin{figure}[h!]
\center{\includegraphics[width=0.3\linewidth]{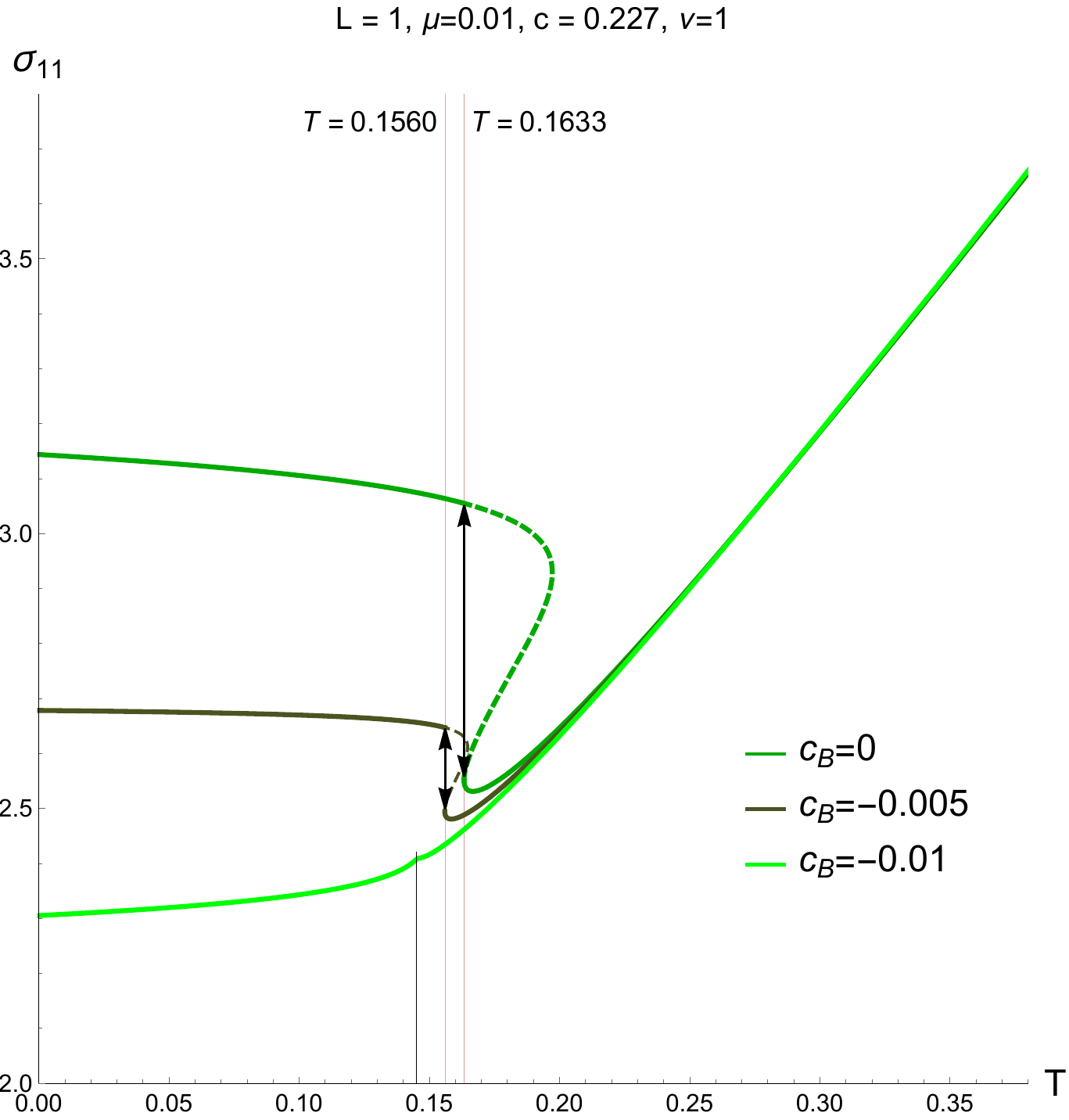}\quad\includegraphics[width=0.3\linewidth]{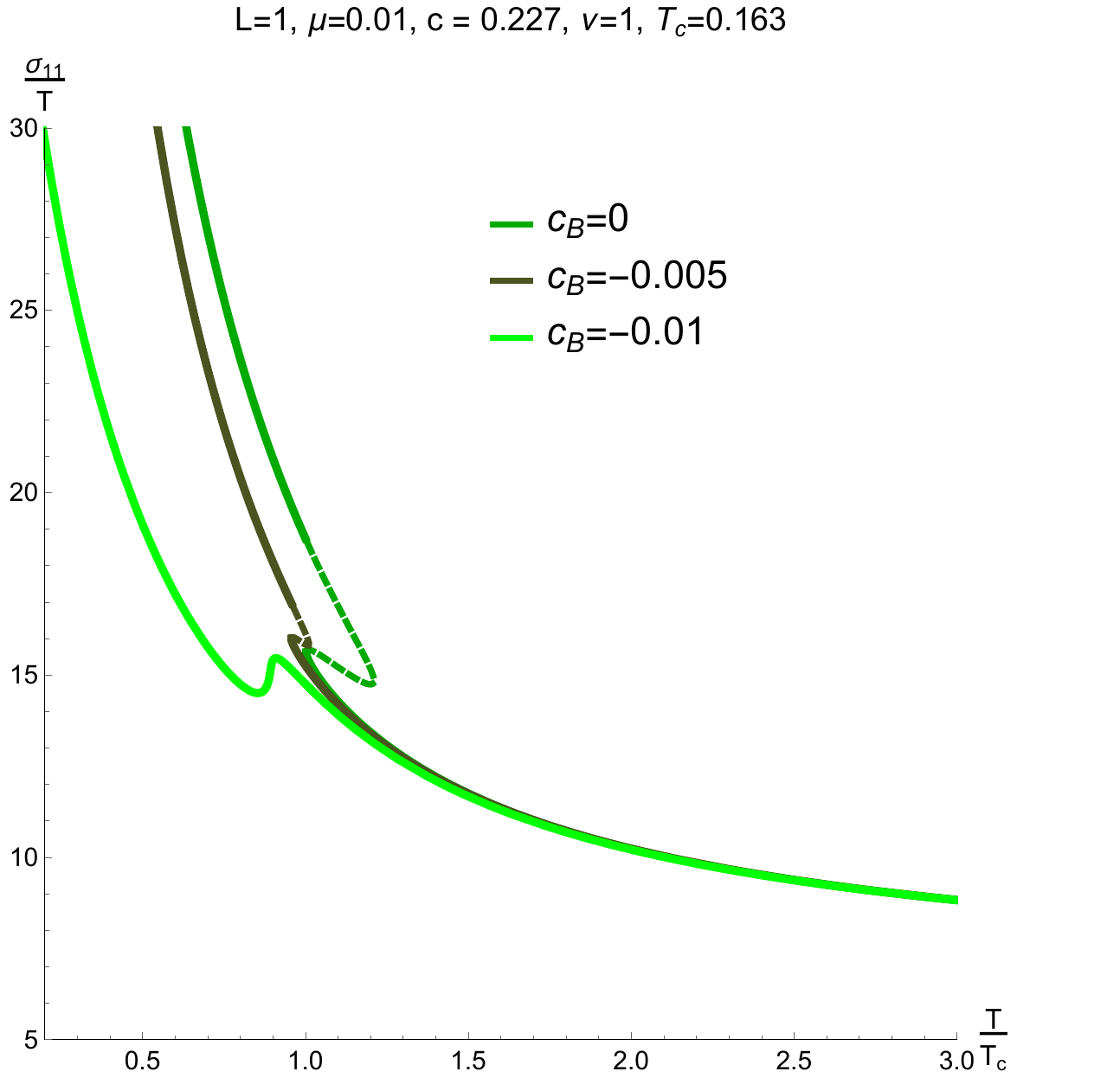} \includegraphics[width=0.3\linewidth]{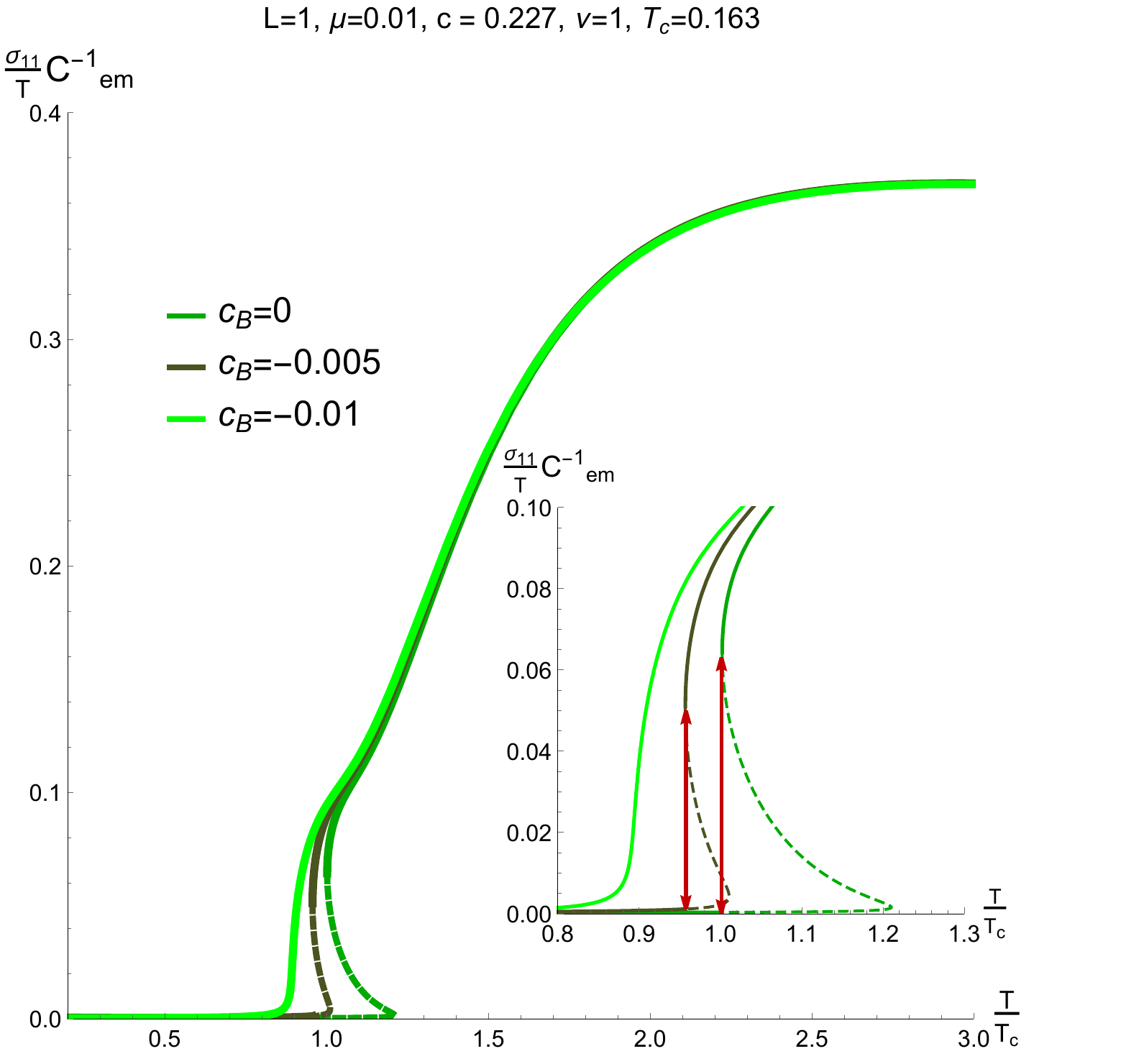}}
\\
A \hspace{40 mm} B  \hspace{40 mm}C
\caption{ A) The dependence of $\sigma^{11}$  on the temperature $T$ is presented with the dilaton coupling $f_0(\phi)=1$. B) The dependence of the ratio 
$\sigma^{11}/T$ on the temperature $T$ is presented with the dilaton coupling $f_0(\phi)=1$. C) The electric conductivity comes with $f_0(\phi)$ function from \eqref{fitfunc}. The plot-in shows the zoom of the main plot near the jumps. }
\label{fig:Fitted}
\end{figure}

In Fig.\ref{fig:Fitted}.A  and Fig.\ref{fig:Fitted}.B the dependence of conductivity $\sigma^{11}$ and ratio $\sigma^{11}/T$ on the temperature $T$ are presented with the gauge kinetic function $f_0(\phi)=1$. In Fig.\ref{fig:Fitted}.C the electric conductivity comes with $f_0(\phi)$ function from \eqref{fitfunc}. The addition of the dilaton coupling changes the behaviour of the conductivity dramatically both below and above the critical value of temperature $T_c$. But some important physical features remain invariant of the kinetic function, i.e. the positions of conductivity's jumps and the dependence on chemical potential and magnetic field. Roughly speaking, the greater the chemical potential or/and magnetic field, the smoother the conductivity. For some values of parameters the unstable phase vanishes. This happens when the temperature as a function of $z_h$ ceases to be three-critical. 
Also, the $\mu$-dependent effects are invariant to the choice of the coupling function.

\newpage

\begin{figure}[h!]
\centering
\includegraphics[width=0.3\linewidth]{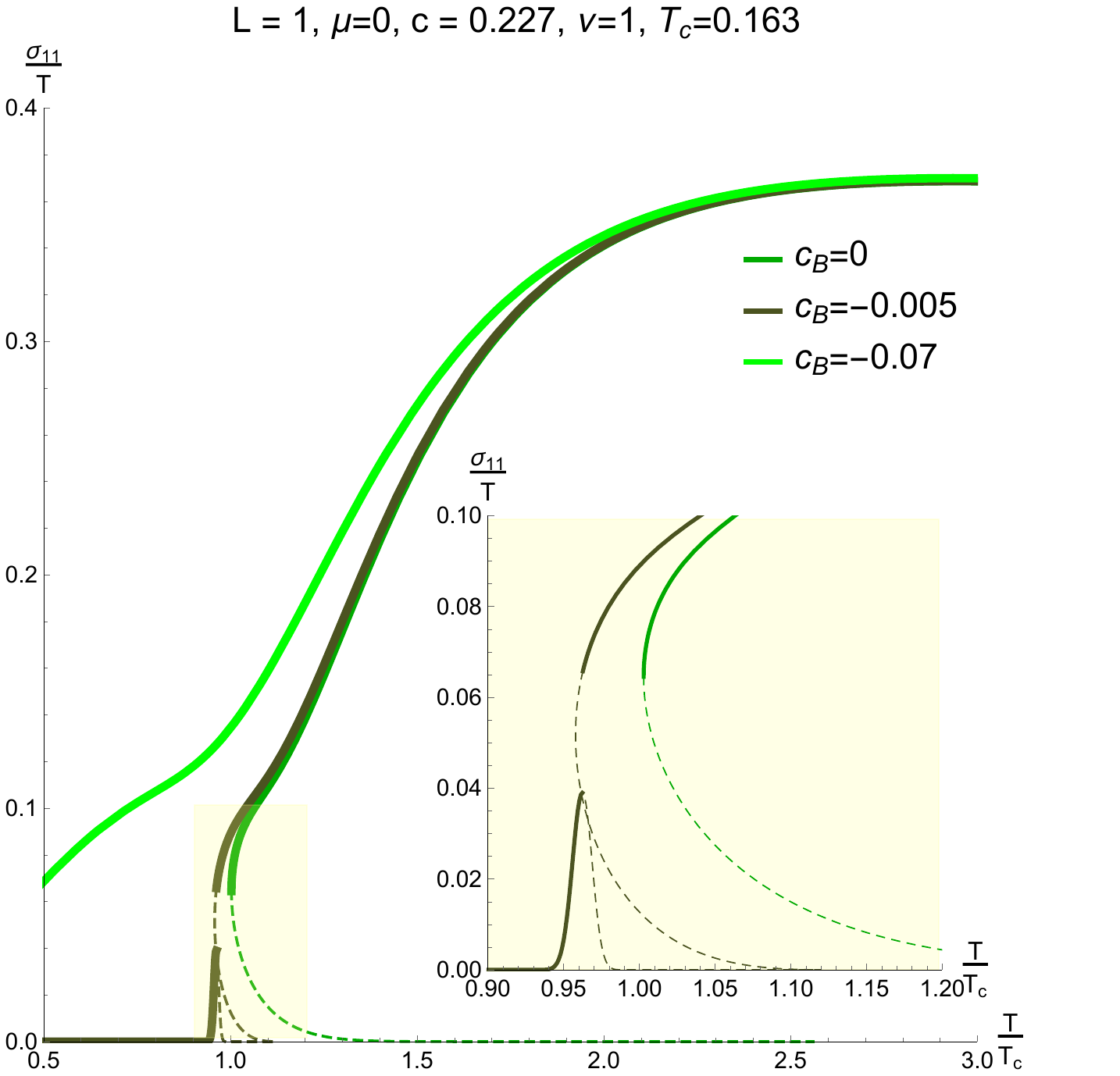} \quad \includegraphics[width=0.3\linewidth]{Plots/M001BN1Fit2.pdf} \quad
\includegraphics[width=0.3\linewidth]{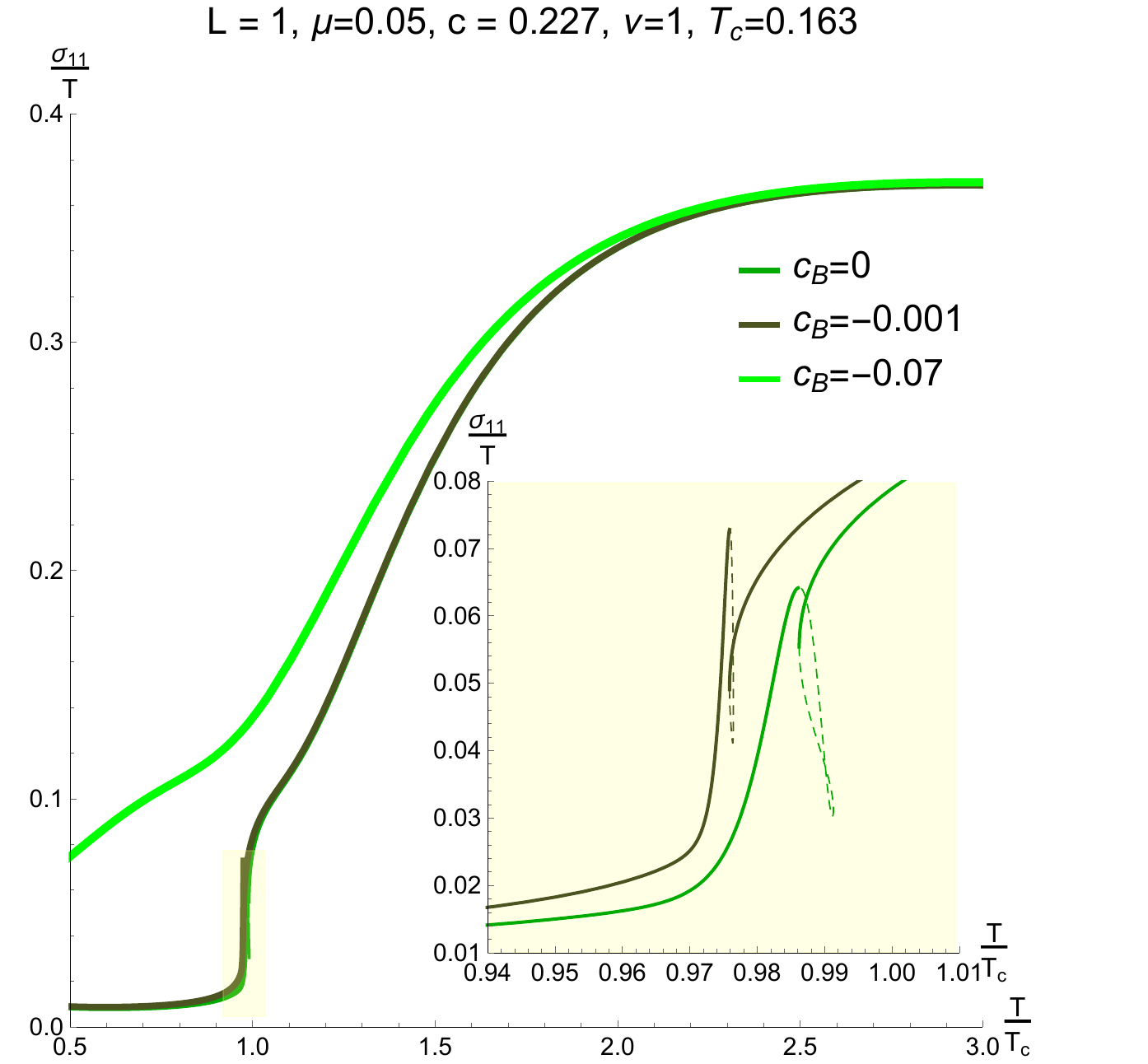}
\\
  A \hspace{50 mm} B \hspace{50 mm} C\\$\,$\\
  \includegraphics[width=0.3\linewidth]{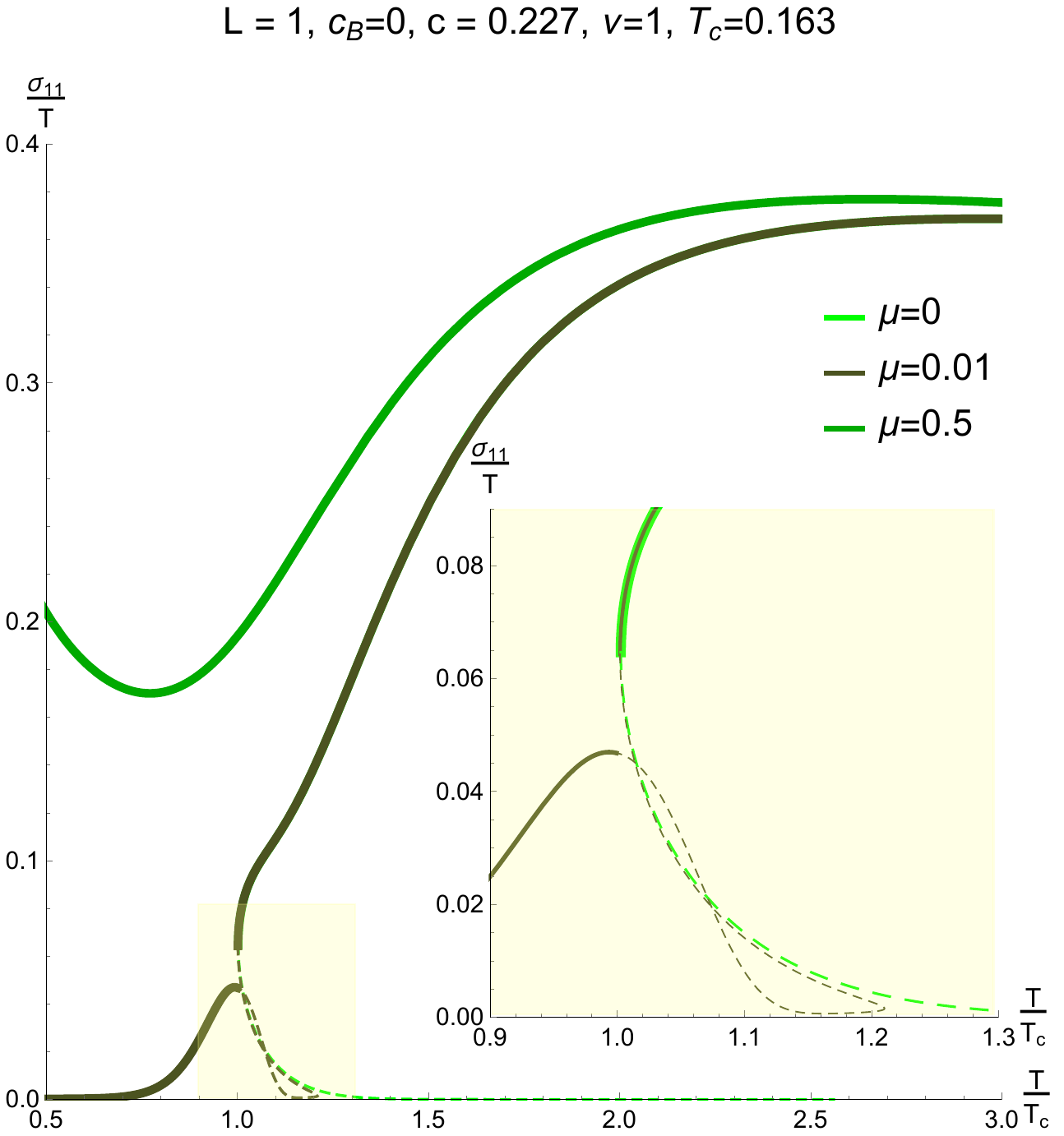} \quad \includegraphics[width=0.3\linewidth]{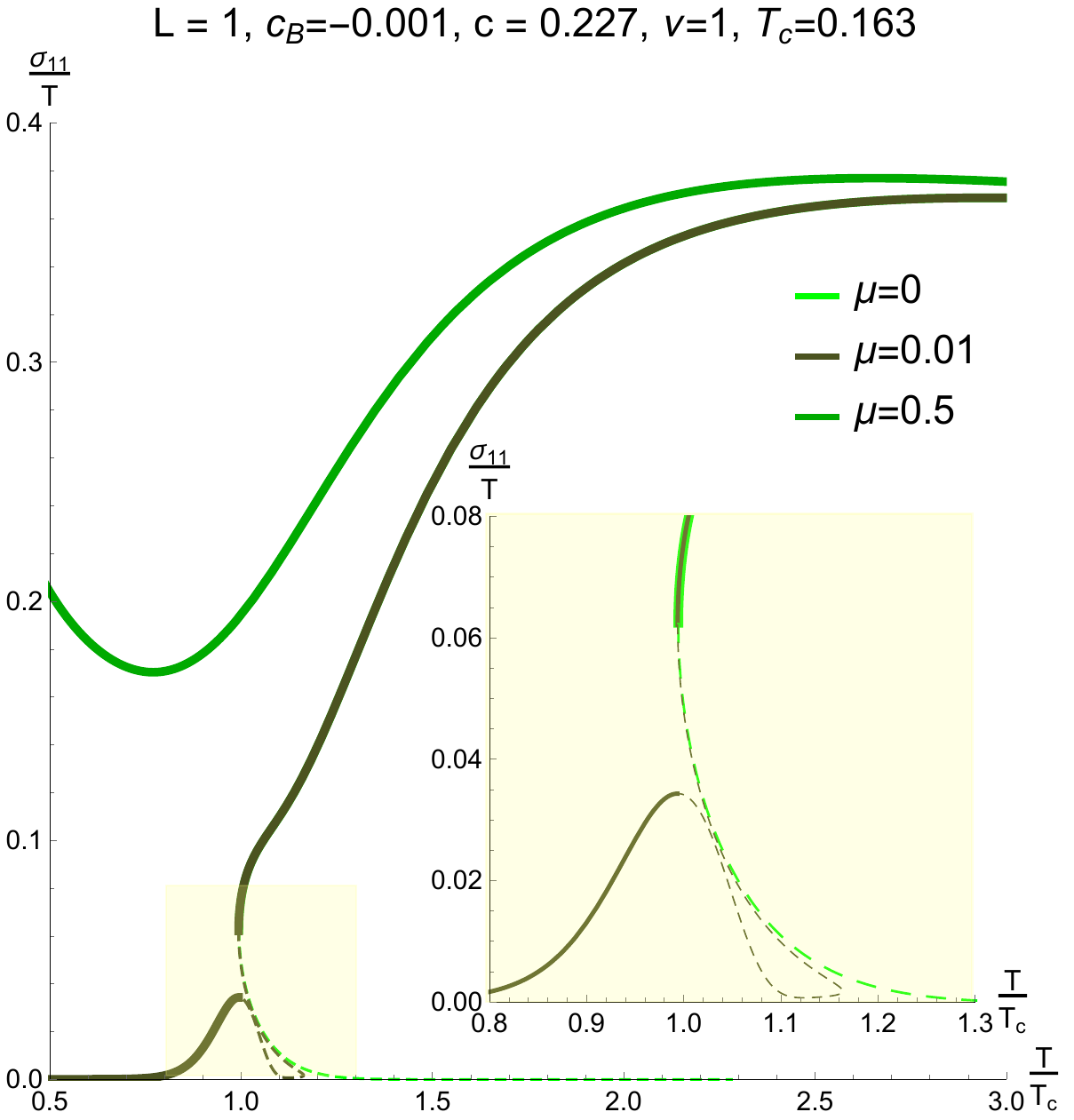} \quad
\includegraphics[width=0.3\linewidth]{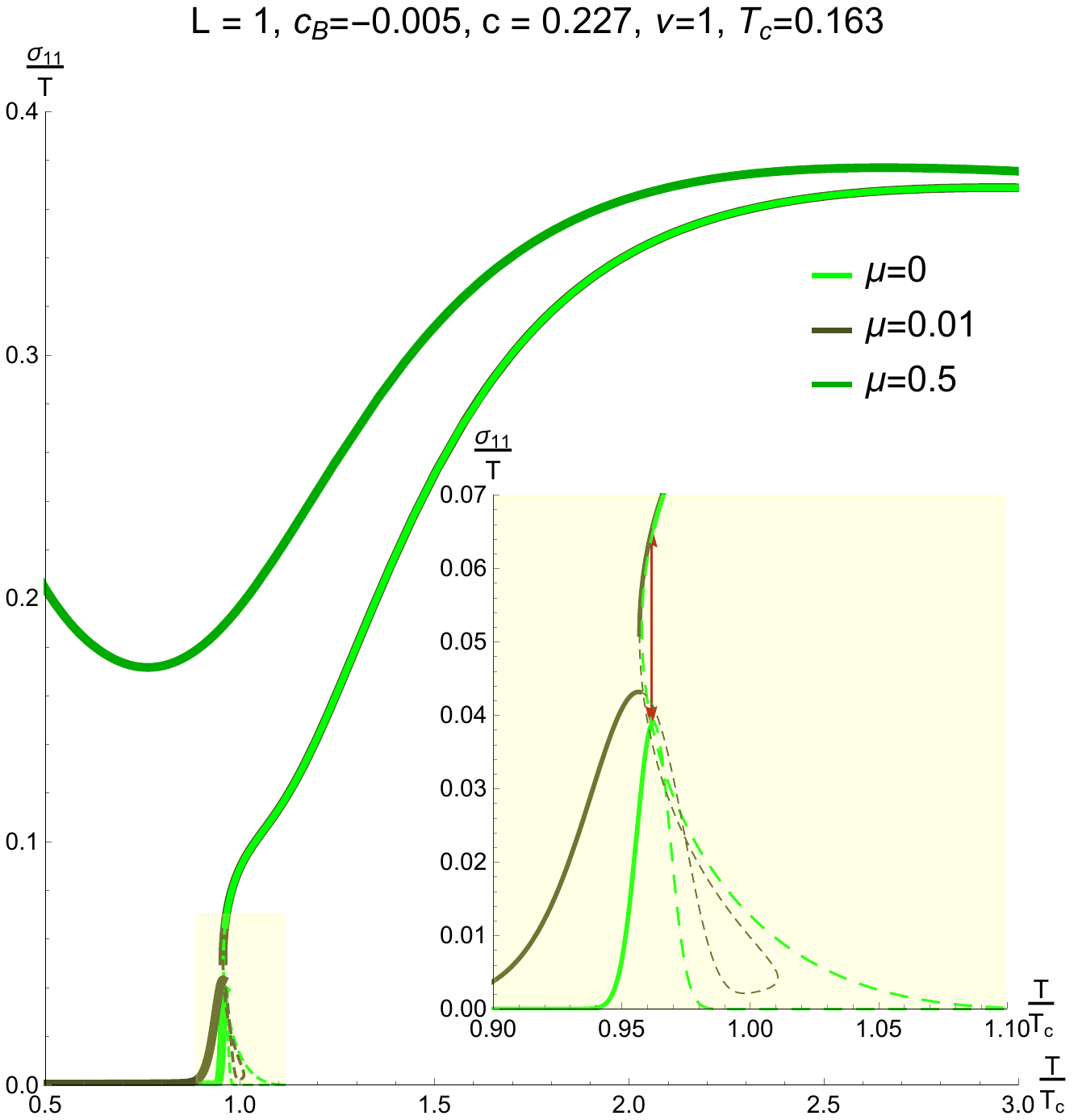}
\\
  D \hspace{50 mm} E \hspace{50 mm} F
\caption{The dependence of $\sigma^{11}/T$ on the normalized temperature $T/T_c$ for different values of magnetic field's  parameter $c_B$ and  $\mu = 0$ (A), $\mu = 0.01$ (B) and $\mu = 0.05$ (C), and for different values of chemical potential $\mu$  and  $c_B = 0$ (D), $c_B =- 0.001$ (E) and $c_B = - 0.005$ (F). Here  $\nu =1$. Dashed lines represent values of $\sigma^{11}$ calculated in thermodynamically unstable phase. The built-in graphs show the zoom of the main plots near the jumps.}
\label{fig:12}
\end{figure}

In Fig.\ref{fig:12} the ratio of electric conductivity to temperature  $\sigma^{11}/T$ on  the normalized temperature $T/T_c$ for $\nu=1$ and  different values of magnetic field's  parameter $c_B$ and  chemical potential $\mu$
 with the coupling function $f_0(\phi)$  given by \eqref{fitfunc}  are presented.
One can see the BB phase transition, which appears at the temperature $T_{BB}(\nu,c_B,\mu)$.  At this temperature, the electric conductivity has a jump. Increasing the chemical potential and/or magnetic field implies vanish of the jump. At ultra-high temperatures, curves on Fig.\ref{fig:12} are approaching some constant value between $0.3$ and $0.31$ that is the same for all parameters. This asymptotic value can be adjusted with an overall factor in $f_0$ function.

\newpage

\begin{figure}[h!]
\centering
\includegraphics[width=0.3\linewidth]{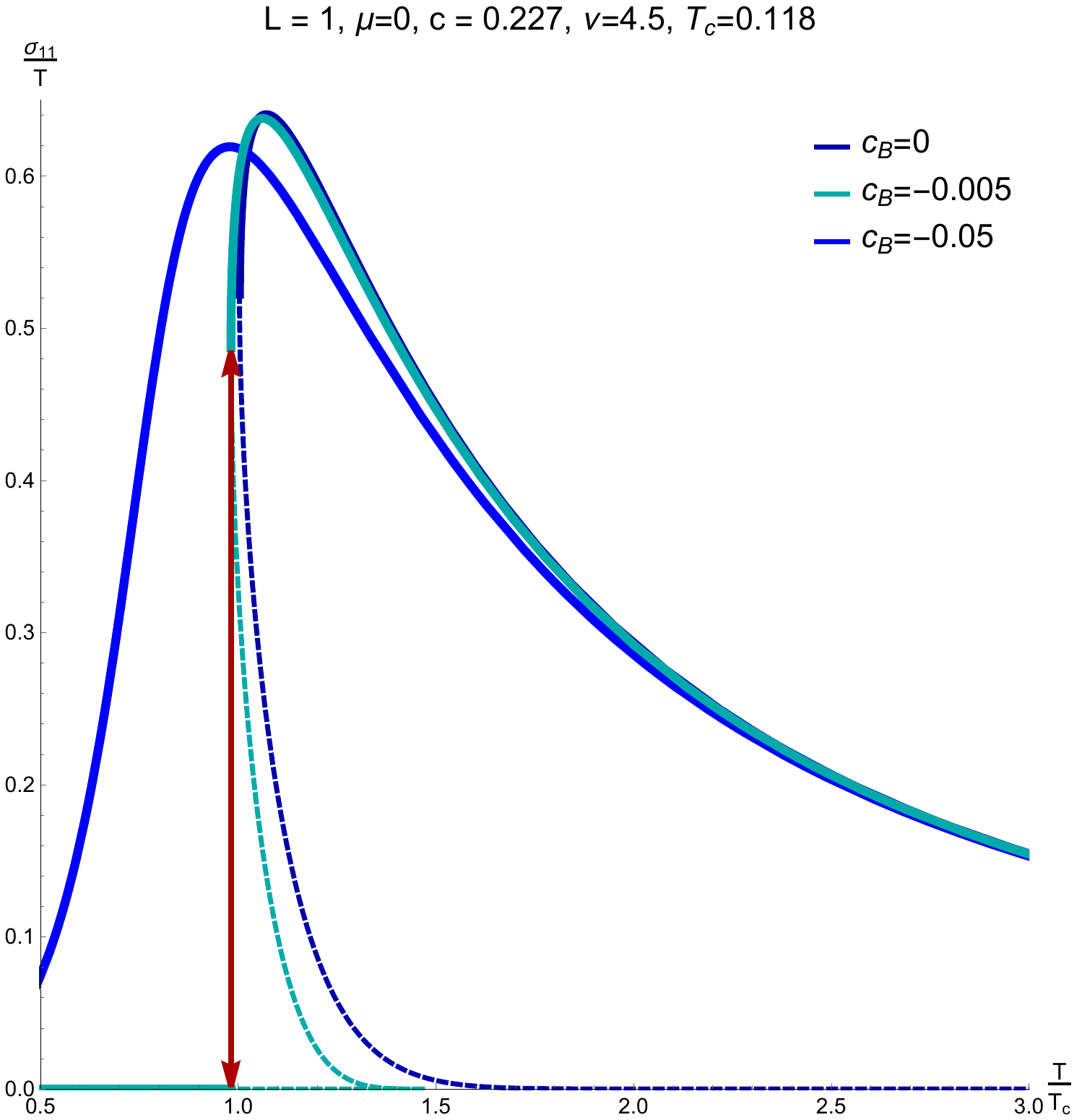} \quad \includegraphics[width=0.3\linewidth]{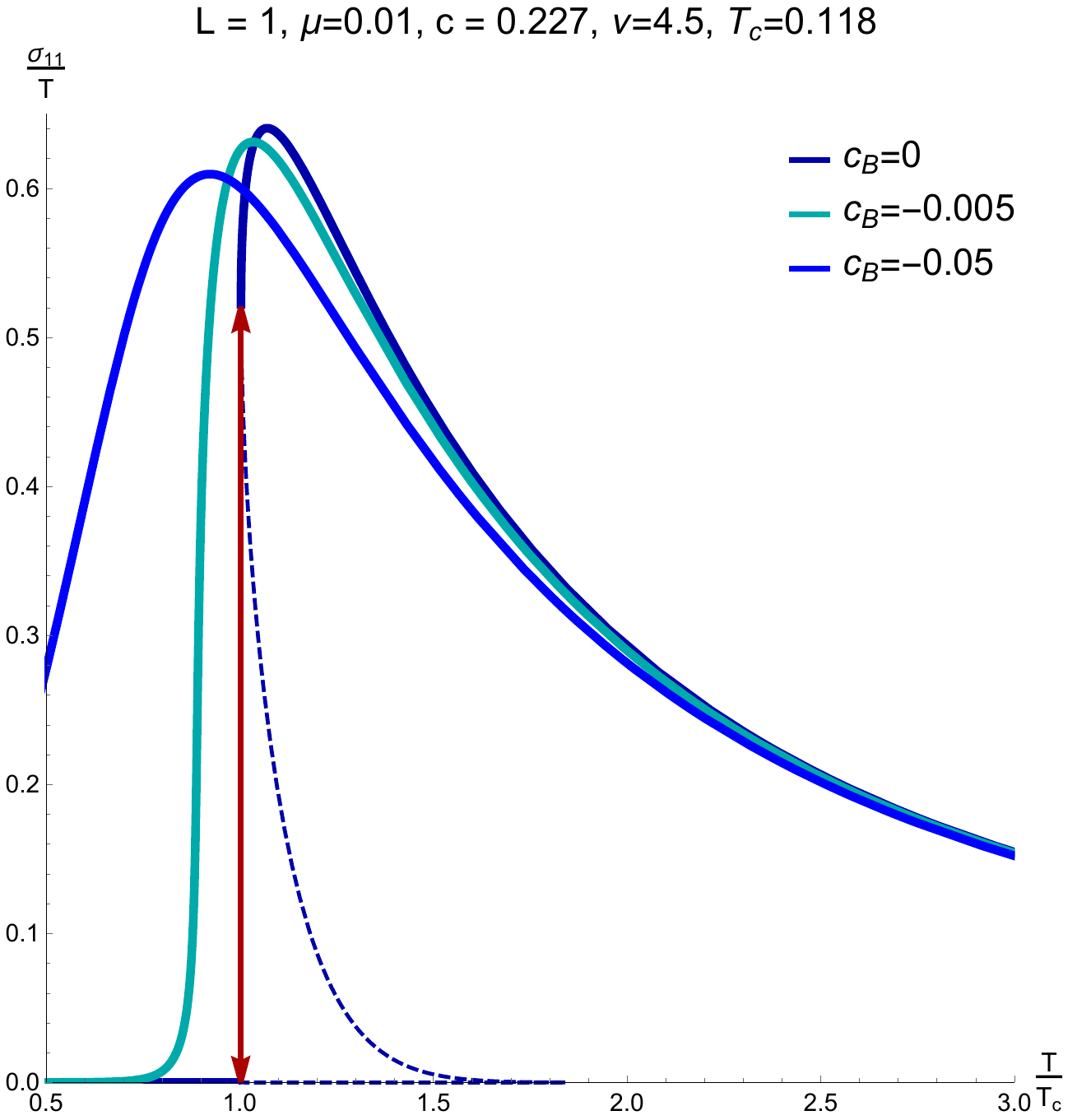} \quad
\includegraphics[width=0.3\linewidth]{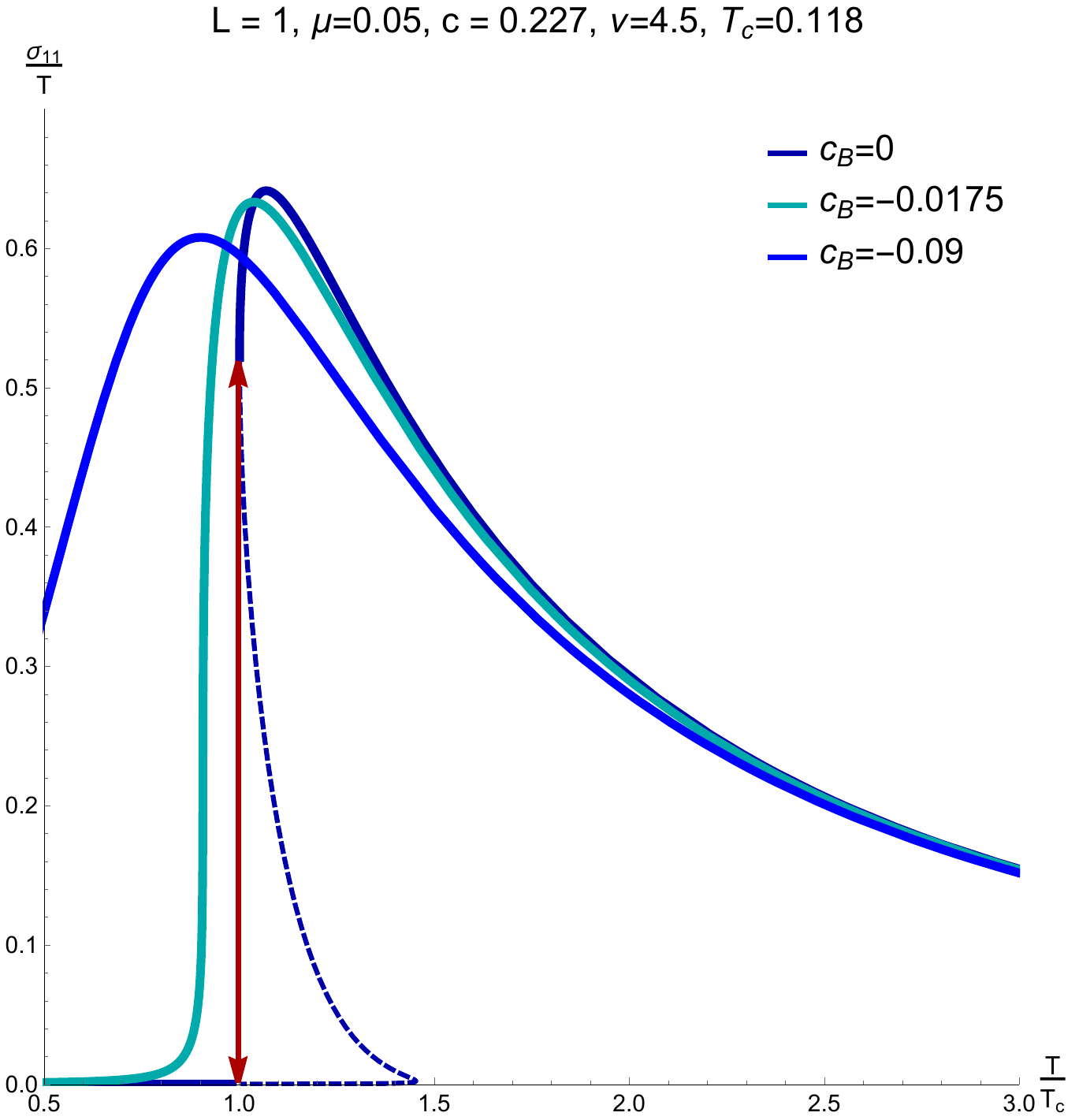}
\\
  A \hspace{50 mm} B \hspace{50 mm} C\\$\,$\\
  \includegraphics[width=0.3\linewidth]{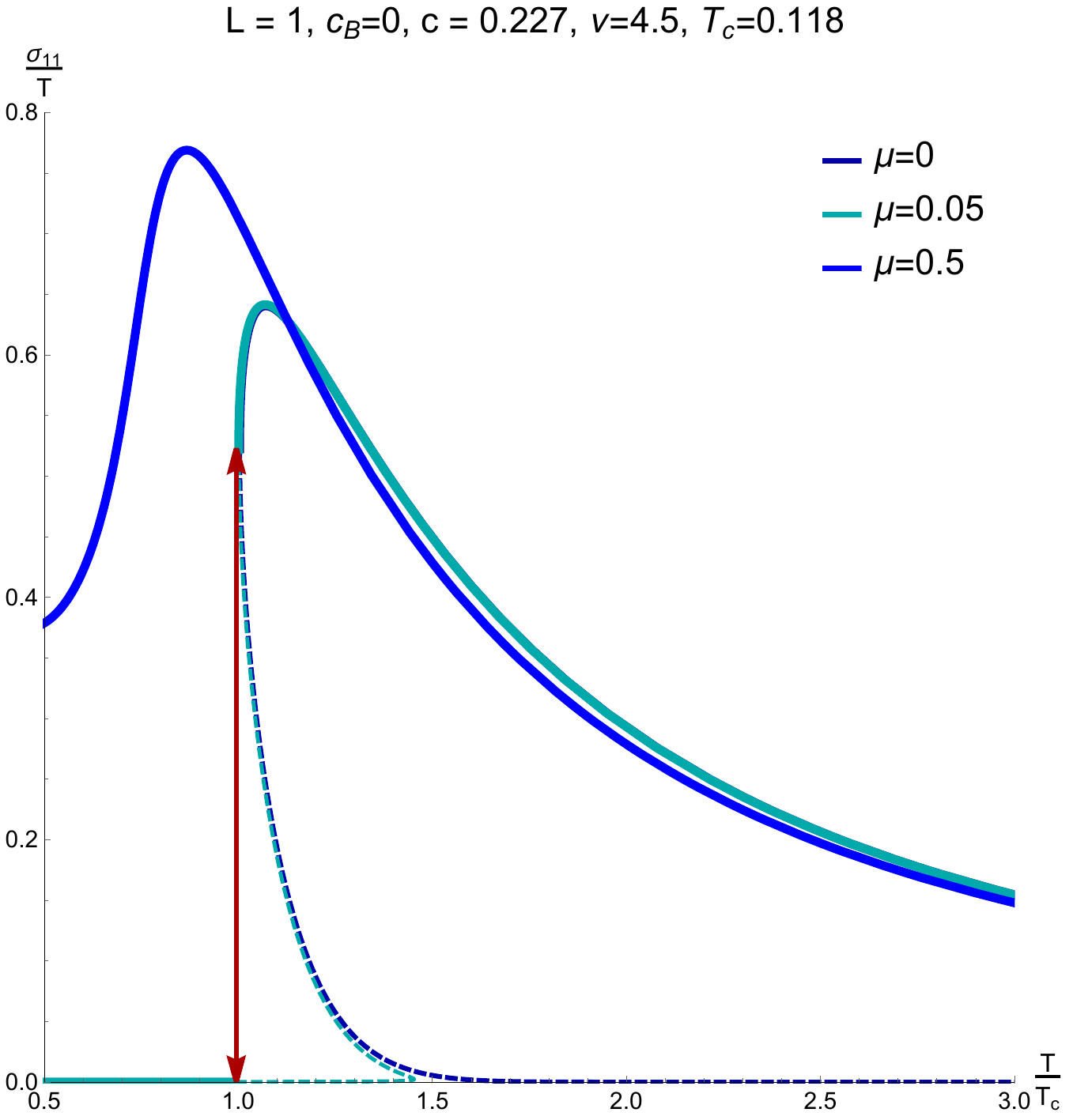} \quad \includegraphics[width=0.3\linewidth]{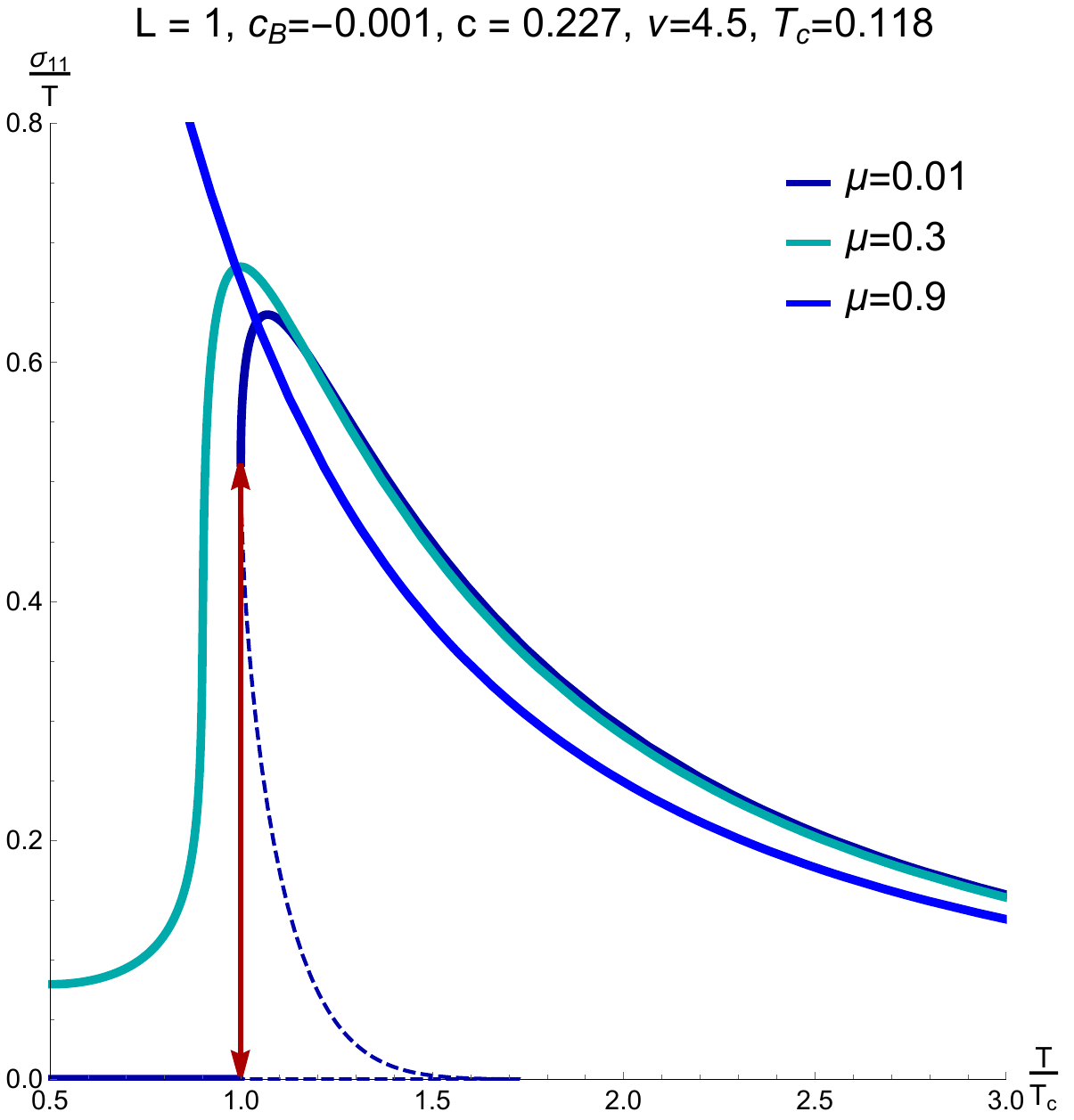} \quad
\includegraphics[width=0.3\linewidth]{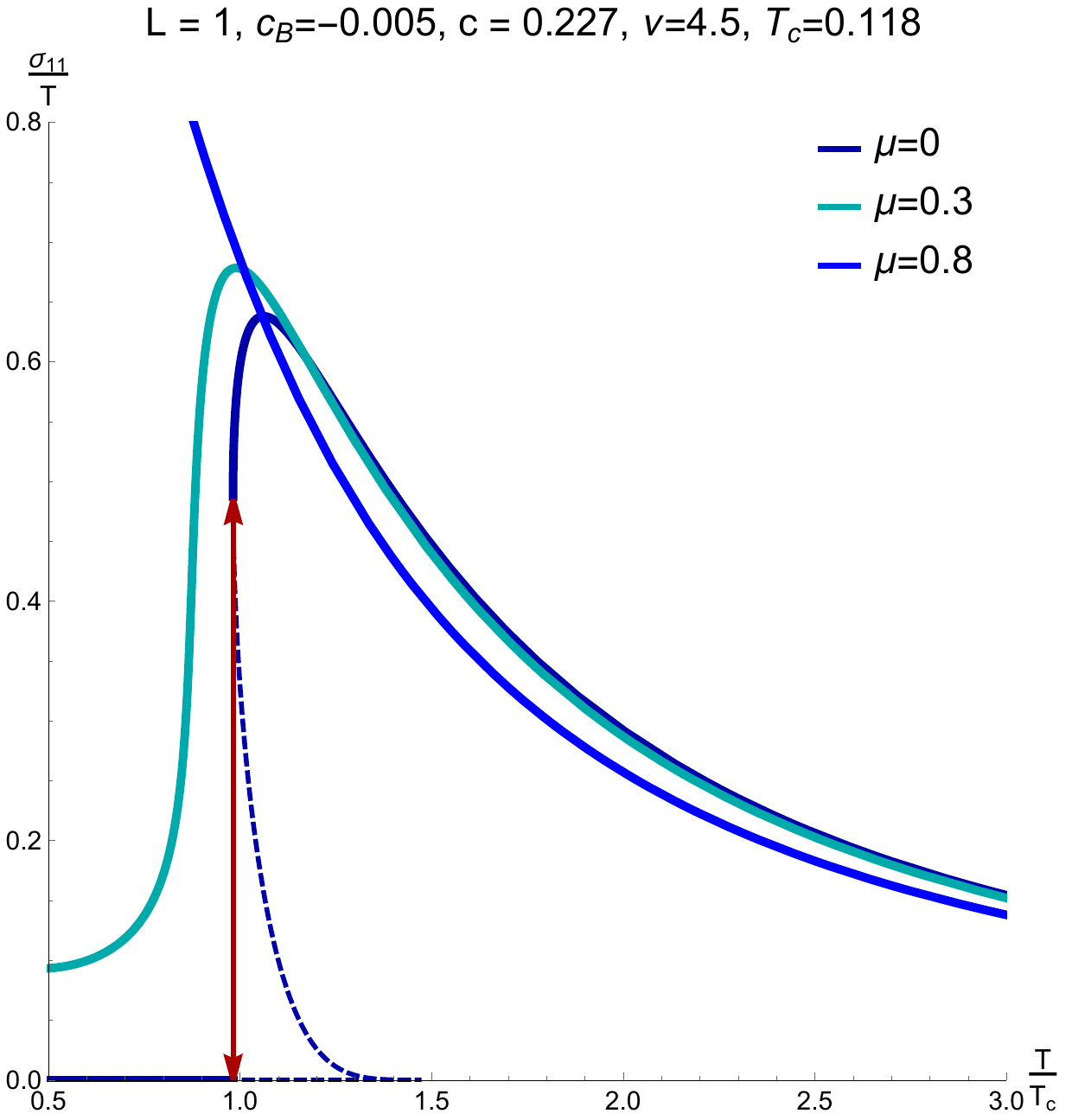}
\\
  D \hspace{50 mm} E \hspace{50 mm} F
\caption{The dependence of $\sigma^{11}/T$ on the normalized temperature $T/T_c$ for different values of magnetic field's  parameter $c_B$ and  $\mu = 0$ (A), $\mu = 0.01$ (B) and $\mu = 0.05$ (C), and for different values of chemical potential $\mu$  and  $c_B = 0$ (D), $c_B =- 0.001$ (E) and $c_B = - 0.005$ (F). Here  $\nu =4.5$. The dashed lines represent values of $\sigma^{11}$ calculated in thermodynamically unstable phase. }
\label{fig:13}
\end{figure}

In Fig.\ref{fig:13} the ratio of electric conductivity to temperature  on  the normalized temperature $T/T_c$ for $\nu=4.5$ and  different values of magnetic field's  parameter $c_B$ and  chemical potential $\mu$
 with the coupling function $f_0(\phi)$  given by \eqref{fitfunc}  are presented. We can see the BB phase transition which appears at the temperature $T_{BB}(\nu,c_B,\mu)$.  At this temperature, the electric conductivity has a jump. Increasing the chemical potential and/or magnetic field implies vanish of the jump. For ultra high values of chemical potential we see that $\sigma^{11}/T$ monotonically decreases. Also note a significant change of behaviour of $\sigma^{11}/T$ for different $\nu$. For $\nu=1$ the curves increase after the phase transition and take some constant value around the SYM one for large temperatures. The opposite happens to $\nu=4.5$ case. The maximum value is reached near the point of phase transition and then the conductivity goes down,  asymptotically approaching zero. This is the consequence of the change in asymptotic behaviour that we have mentioned in \eqref{sigma_11_asymp}. The plot Fig.\ref{fig3} gives more detailed information about this phenomenon.  The DC conductivity becomes so small for large $T$, that the QGP is almost opaque along the heavy-ions collision line.

\subsection{Calculation of $\sigma^{33}$}
\begin{figure}[h!]
\centering
\includegraphics[width=0.3\linewidth]{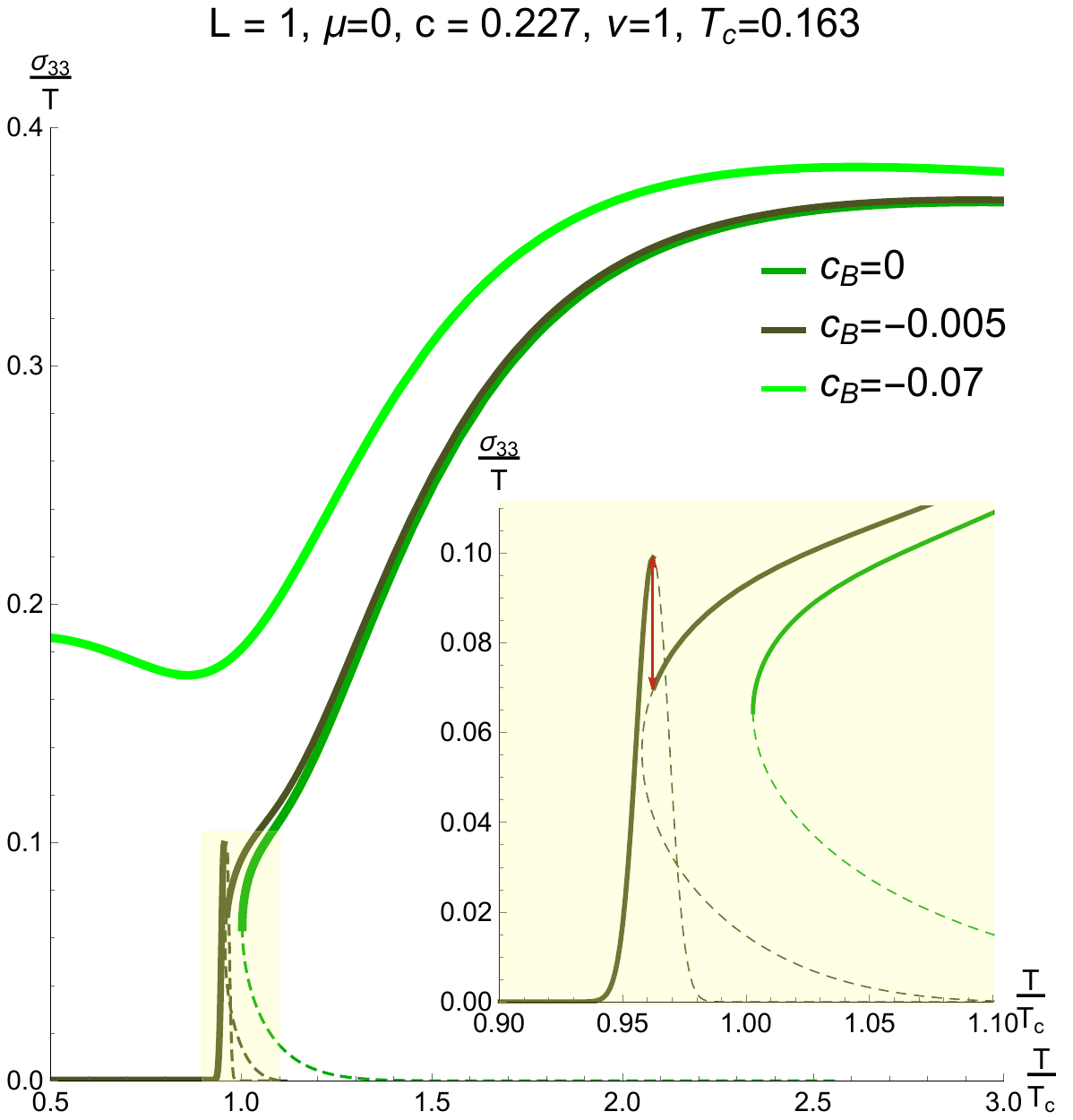} \quad \includegraphics[width=0.3\linewidth]{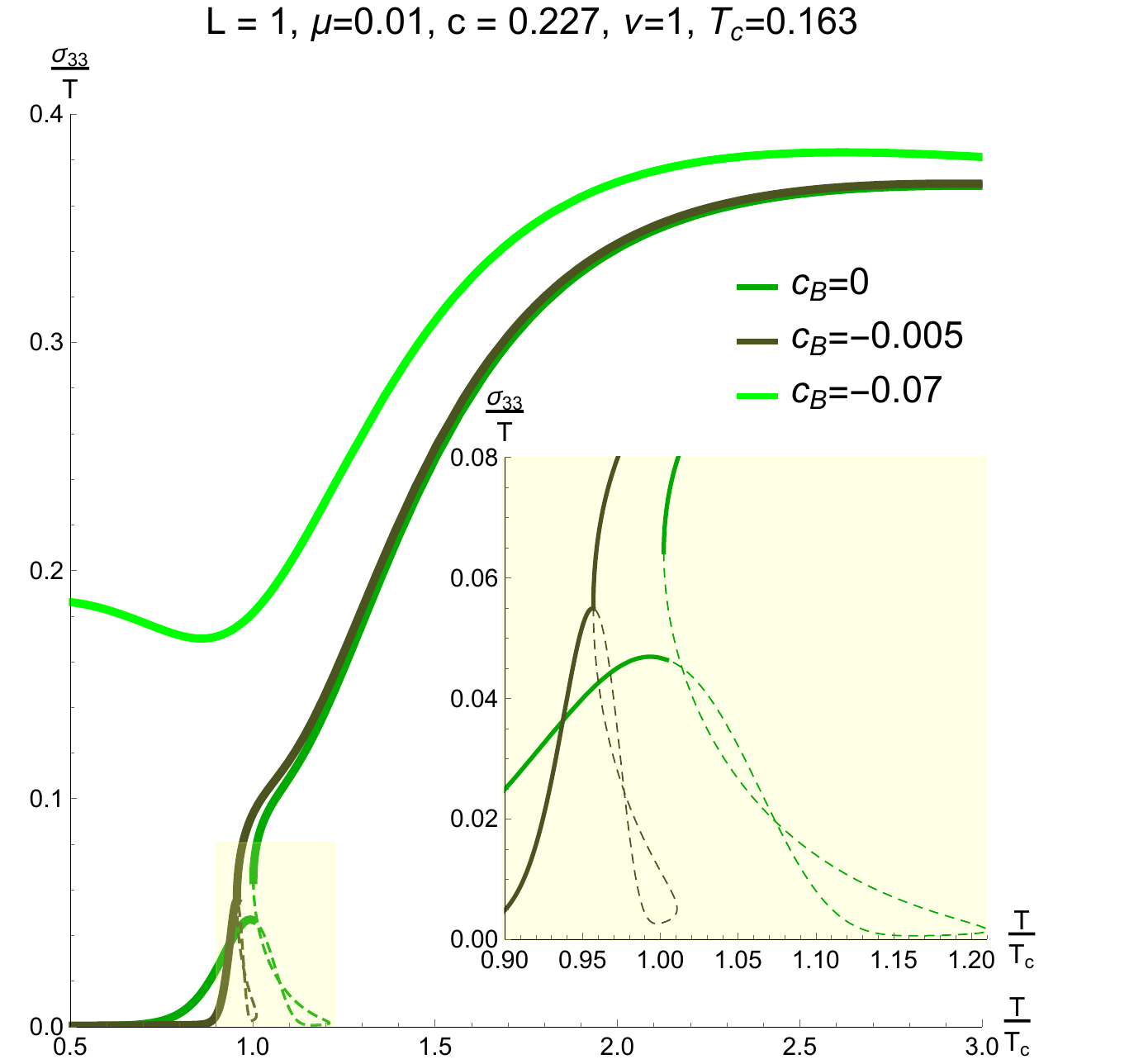} \quad
\includegraphics[width=0.3\linewidth]{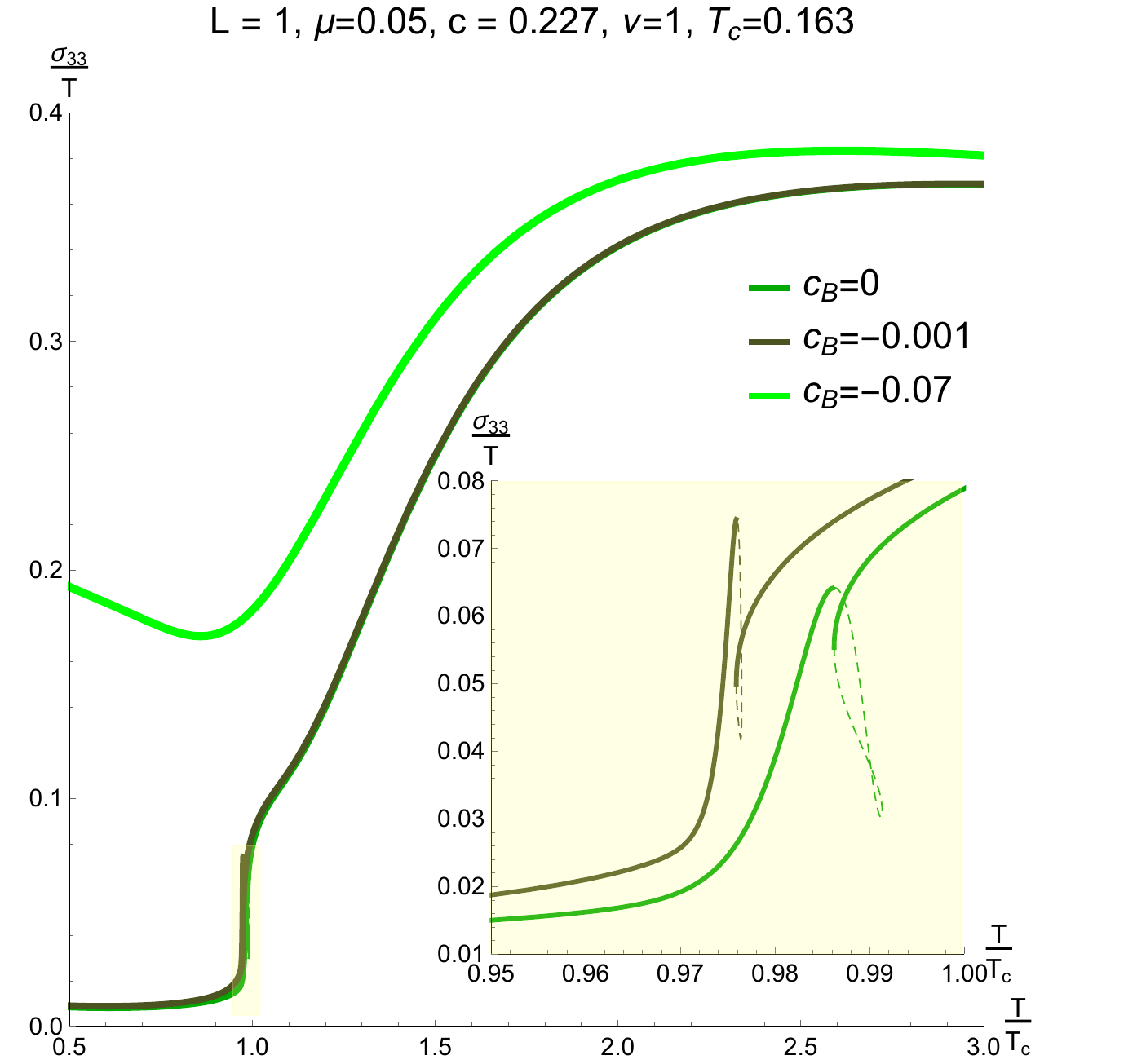}
\\
  A \hspace{50 mm} B \hspace{50 mm} C\\$\,$\\
  \includegraphics[width=0.3\linewidth]{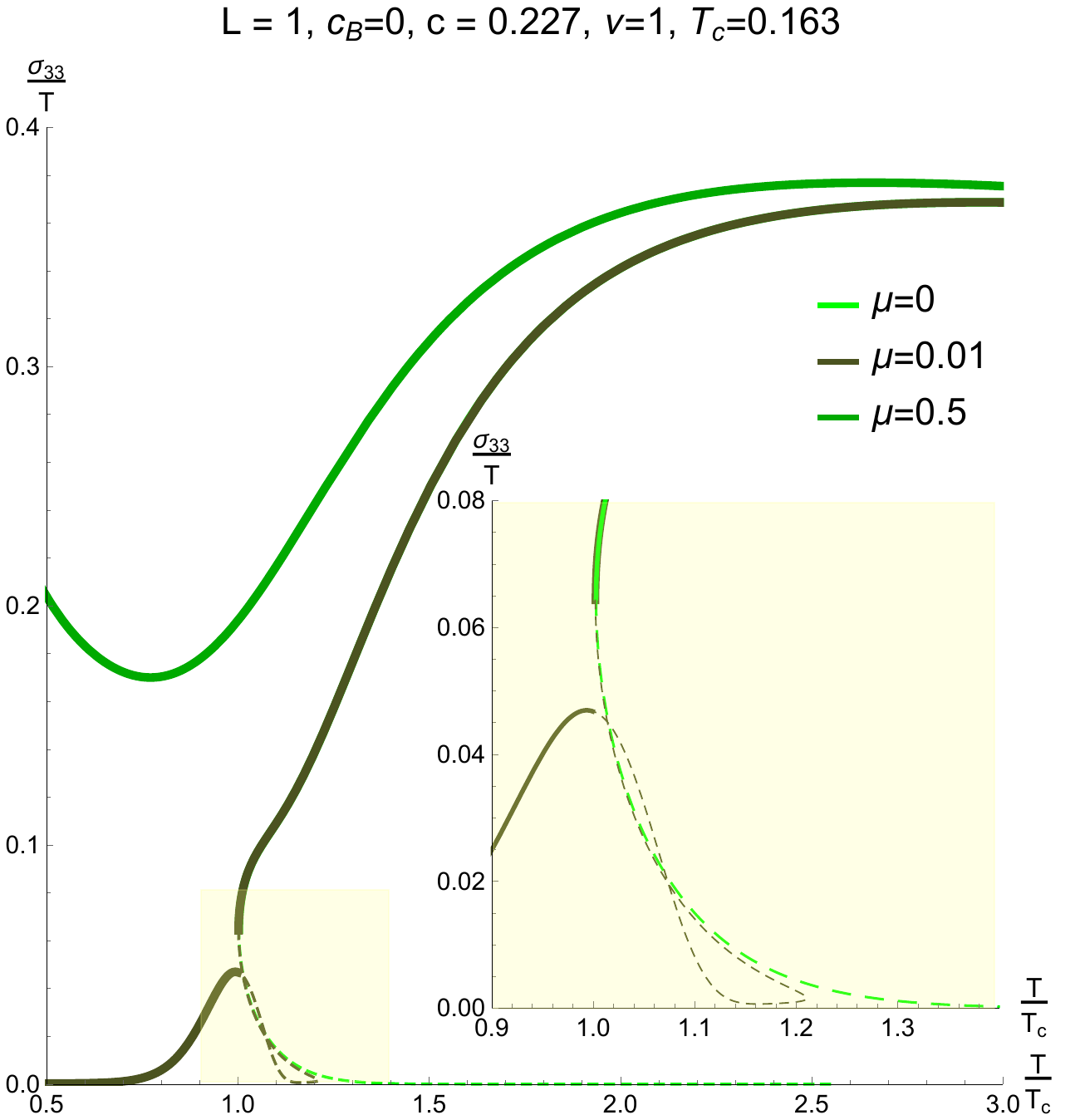} \quad \includegraphics[width=0.3\linewidth]{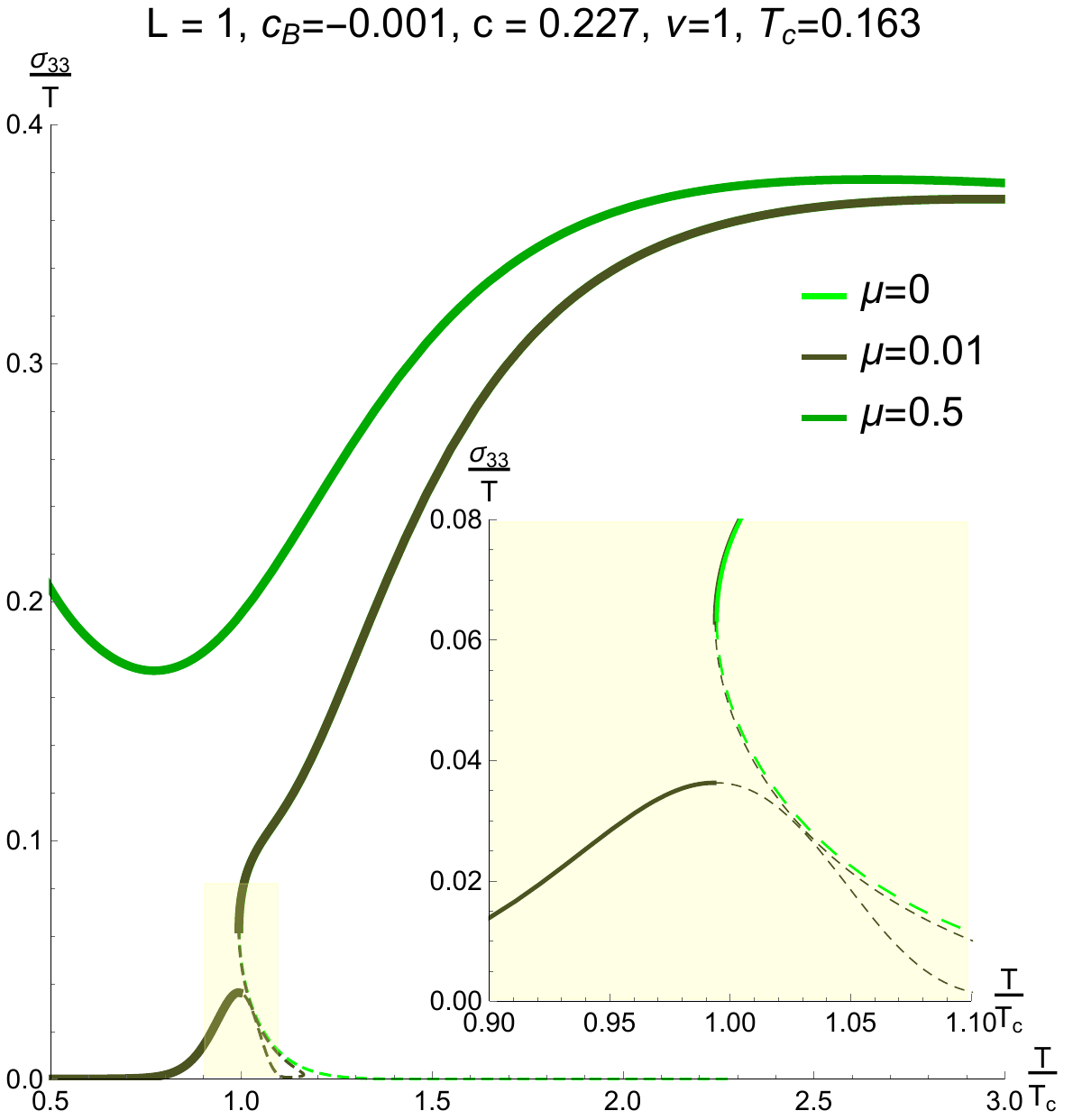} \quad
\includegraphics[width=0.3\linewidth]{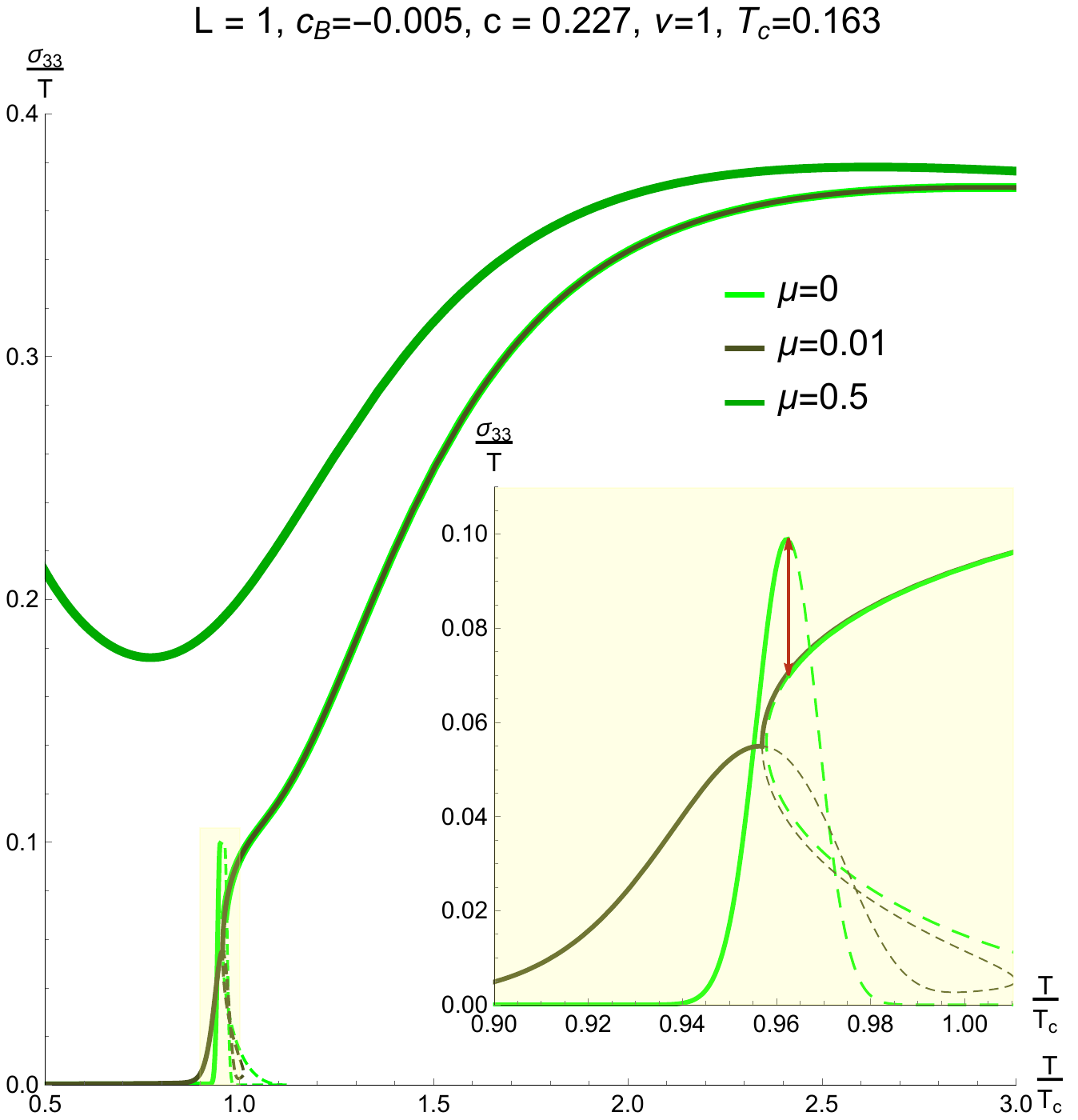}
\\
  D \hspace{50 mm} E \hspace{50 mm} F
\caption{The dependence of $\sigma^{33}/T$ on the normalized temperature $T/T_c$ for different values of magnetic field's  parameter $c_B$ and  $\mu = 0$ (A), $\mu = 0.01$ (B) and $\mu = 0.05$ (C), and for different values of chemical potential $\mu$  and  $c_B = 0$ (D), $c_B =- 0.001$ (E) and $c_B = - 0.005$ (F). Here  $\nu =1$. Dashed lines represent values of $\sigma^{11}$ calculated in thermodynamically unstable phase. The built-in graphs show the zoom of the main plots near the jumps.}
\label{fig:14}
\end{figure}

In Fig.\ref{fig:14} the ratio of electric conductivity to temperature  $\sigma^{33}/T$ on  the normalized temperature $T/T_c$ for $\nu=1$ and  different values of magnetic field's  parameter $c_B$ and  chemical potential $\mu$
 with the coupling function $f_0(\phi)$  given by \eqref{fitfunc}  are presented.
We can see the BB phase transition which appears at the temperature $T_{BB}(\nu,c_B,\mu)$.  At this temperature the electric conductivity has a jump. Increasing the chemical potential and/or magnetic field implies vanish of the jump.
\newpage

\begin{figure}[h!]
\centering
\includegraphics[width=0.3\linewidth]{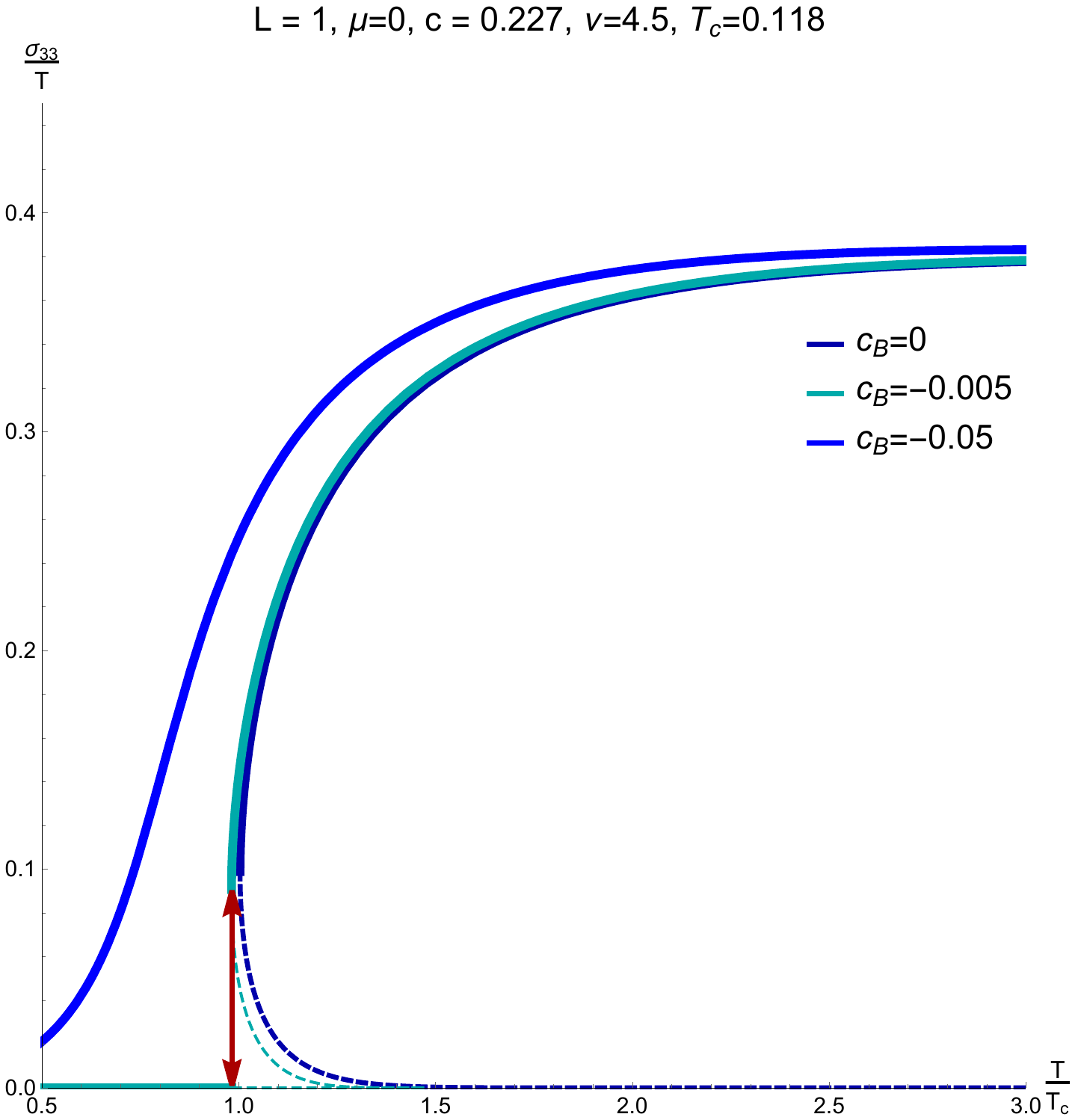} \quad \includegraphics[width=0.3\linewidth]{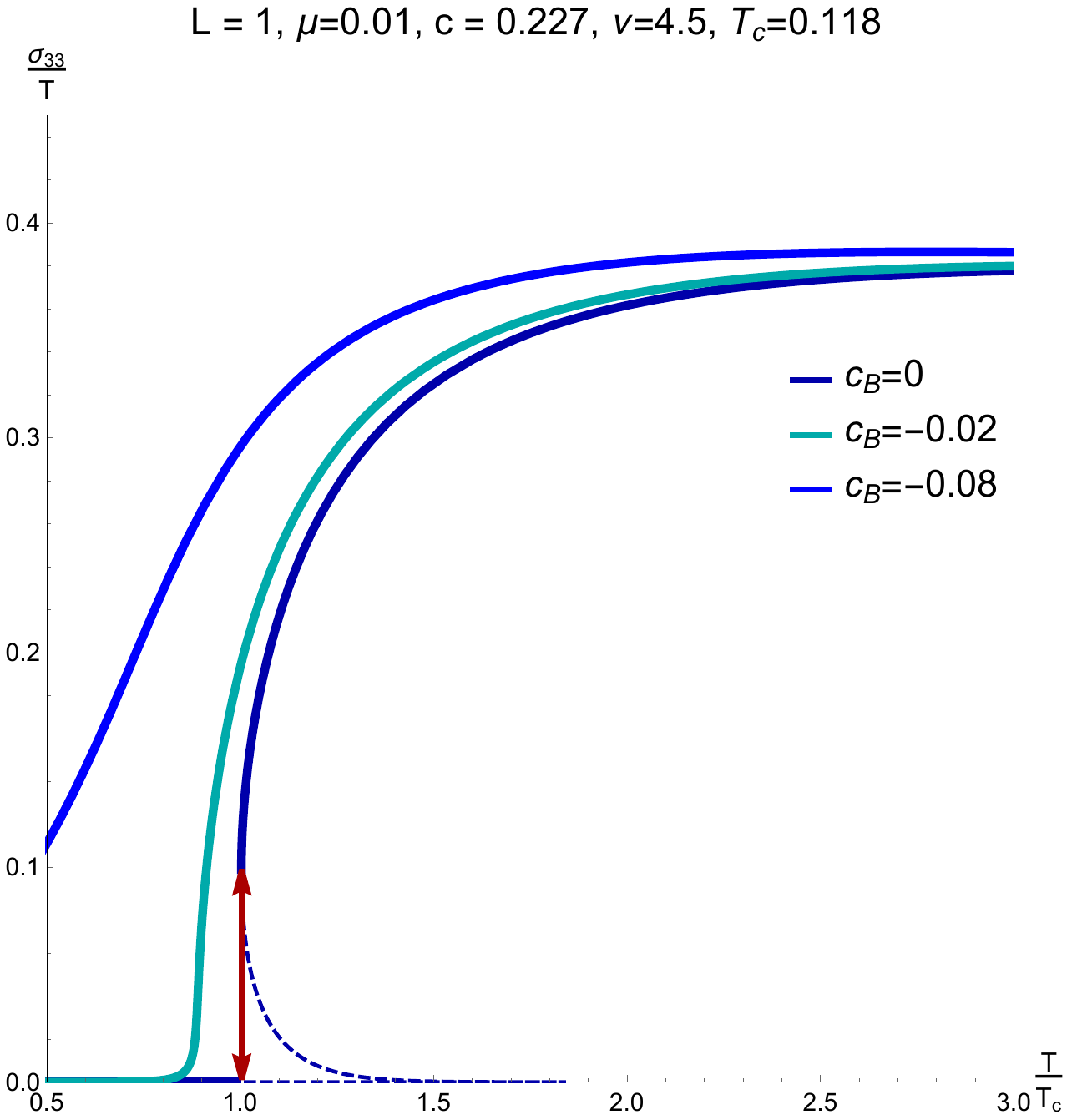} \quad
\includegraphics[width=0.3\linewidth]{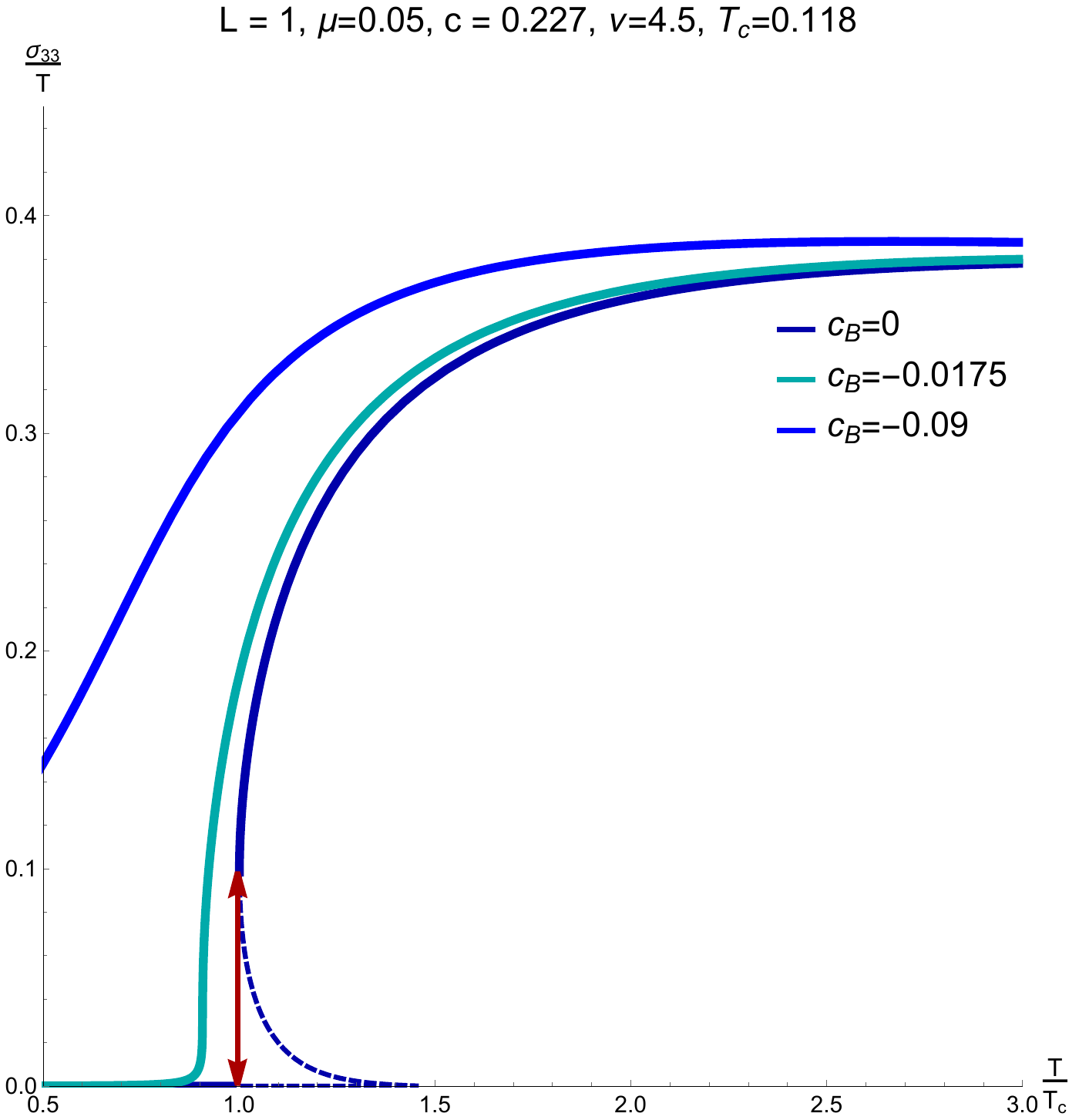}
\\
  A \hspace{50 mm} B \hspace{50 mm} C\\$\,$\\
  \includegraphics[width=0.3\linewidth]{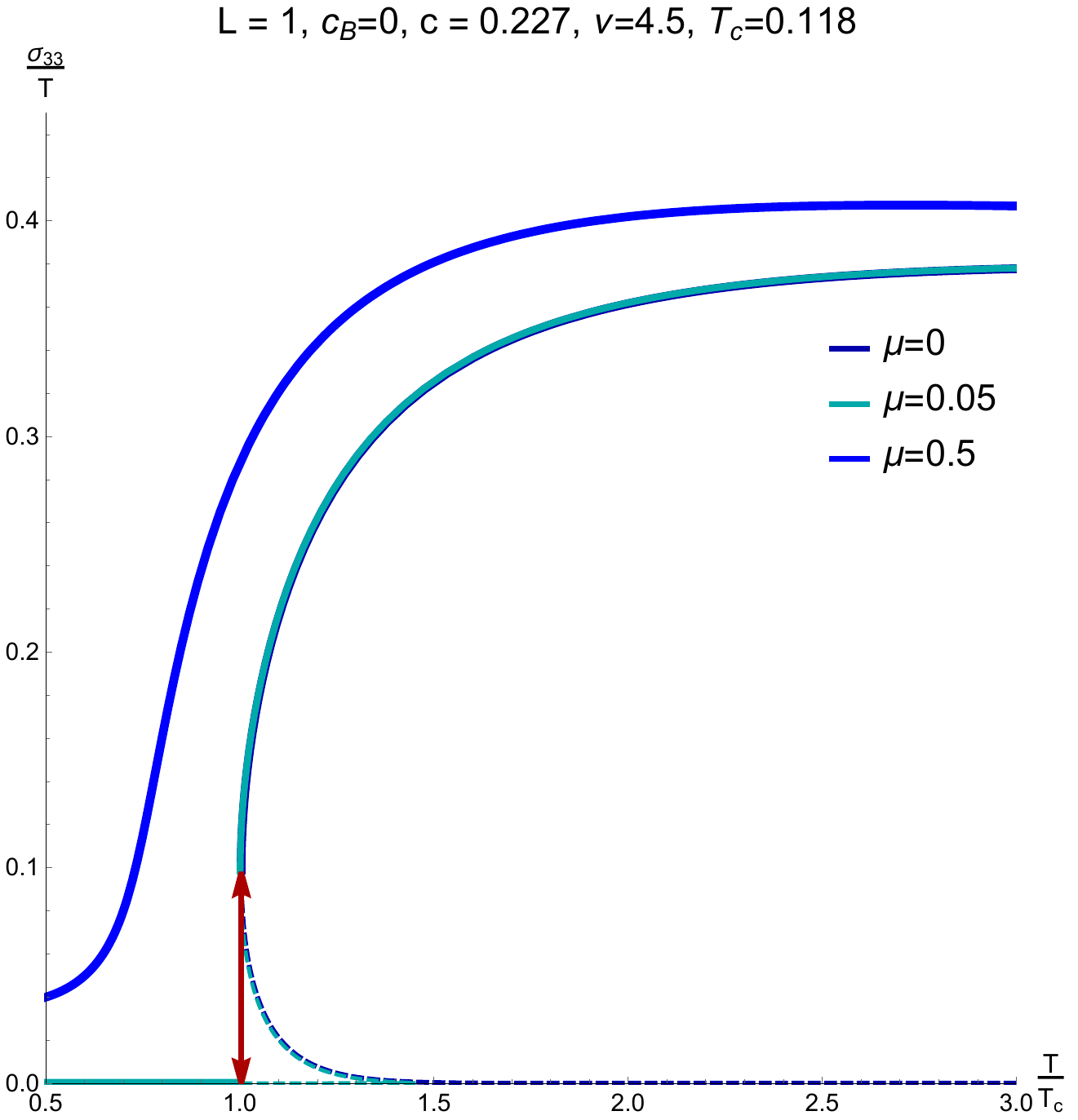} \quad \includegraphics[width=0.3\linewidth]{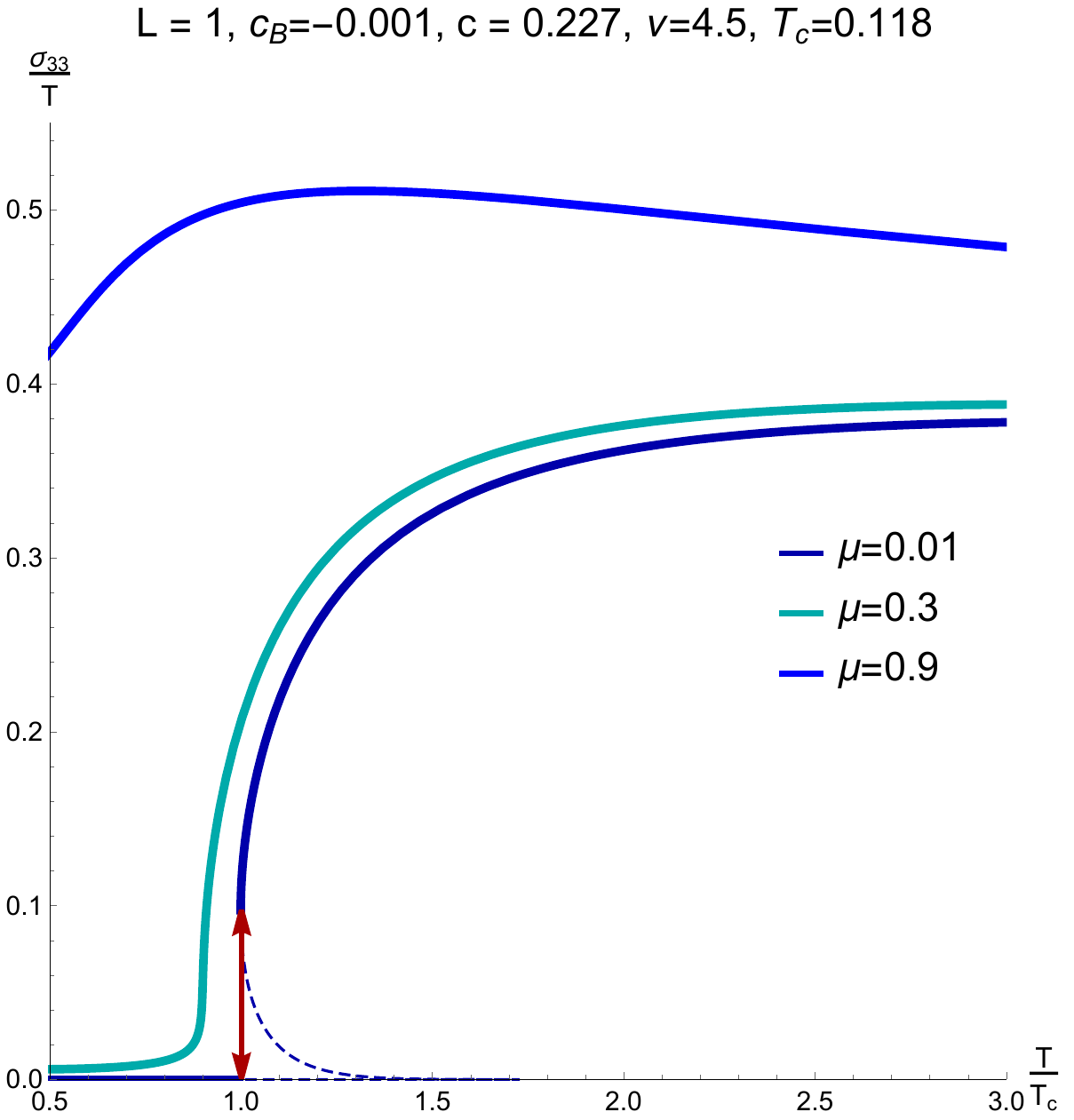} \quad
\includegraphics[width=0.3\linewidth]{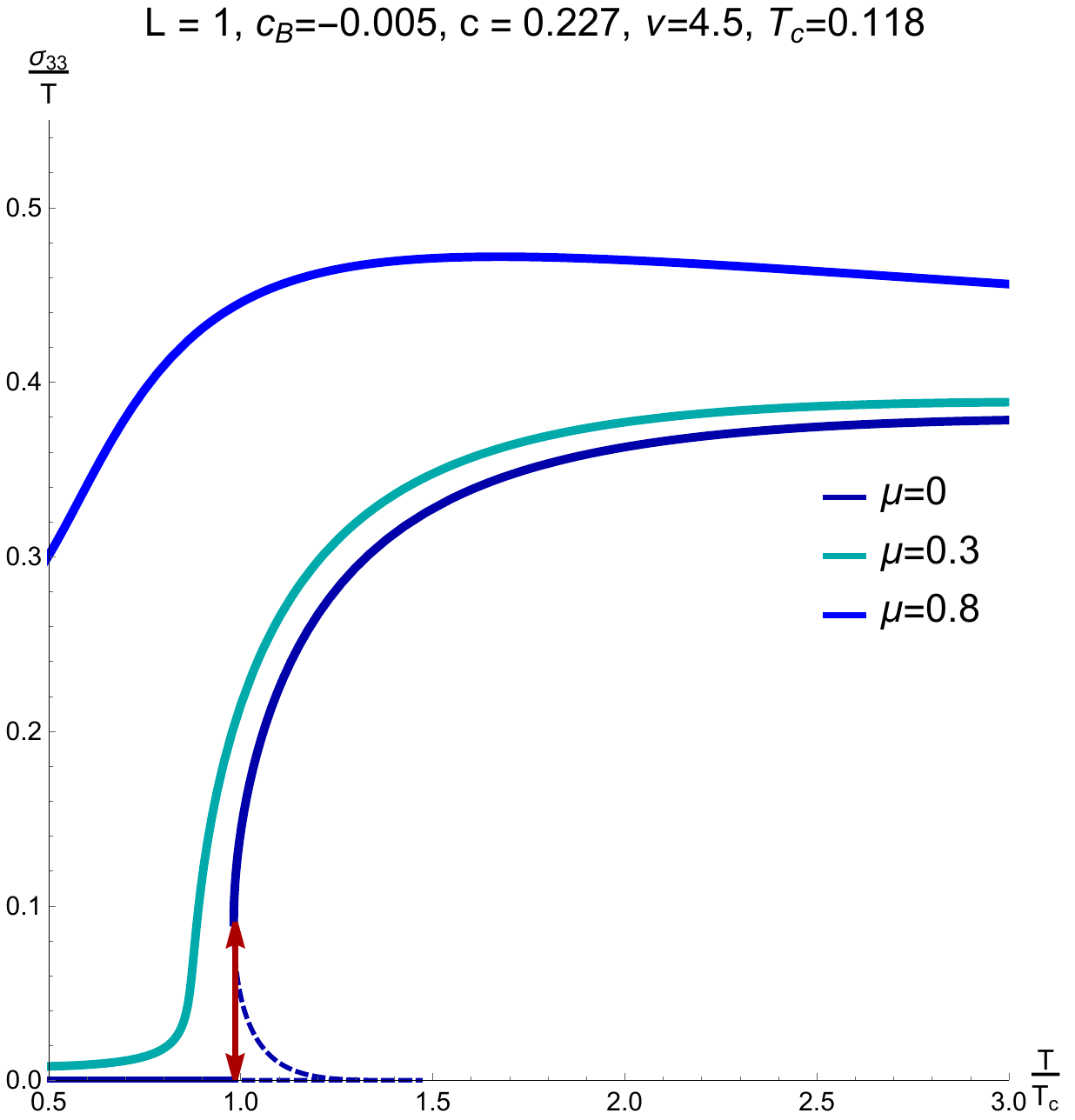}
\\
  D \hspace{50 mm} E \hspace{50 mm} F
\caption{The dependence of $\sigma^{33}/T$ on the normalized temperature $T/T_c$ for different values of magnetic field's  parameter $c_B$ and  $\mu = 0$ (A), $\mu = 0.01$ (B) and $\mu = 0.05$ (C), and for different values of chemical potential $\mu$  and  $c_B = 0$ (D), $c_B =- 0.001$ (E) and $c_B = - 0.005$ (F). Here  $\nu =4.5$. Dashed lines represent values of $\sigma^{33}$ calculated in thermodynamically unstable phase.}
\label{fig:15}
\end{figure}

In Fig.\ref{fig:15} the ratio of electric conductivity to temperature  $\sigma^{33}/T$ on  the normalized temperature $T/T_c$ for $\nu=4.5$ and  different values of magnetic field's  parameter $c_B$ and  chemical potential $\mu$
 with the dilaton coupling function $f_0(\phi)$  given by \eqref{fitfunc}  are presented.
  We can see the BB phase transition which appears at the temperature $T_{BB}(\nu,c_B,\mu)$.  At this temperature the electric conductivity has a jump. Increasing the chemical potential and/or magnetic field implies vanish of the jump. 

\newpage
\subsection{Comparison  of $\sigma^{22}$ and $\sigma^{33}$  }\label{sec:sigma22}

\begin{figure}[h!]
\center{\includegraphics[width=0.3\linewidth]{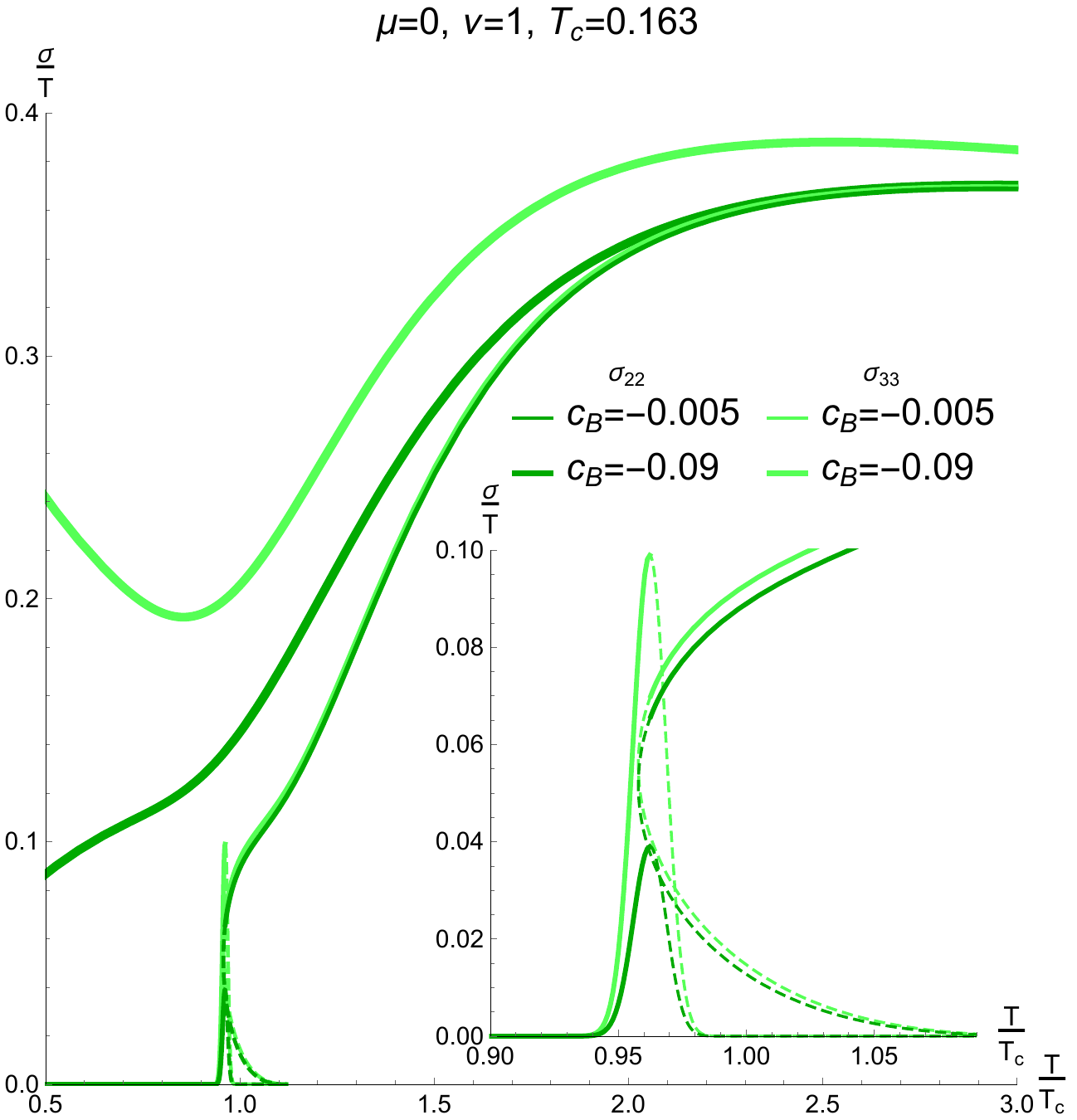} \quad \includegraphics[width=0.3\linewidth]{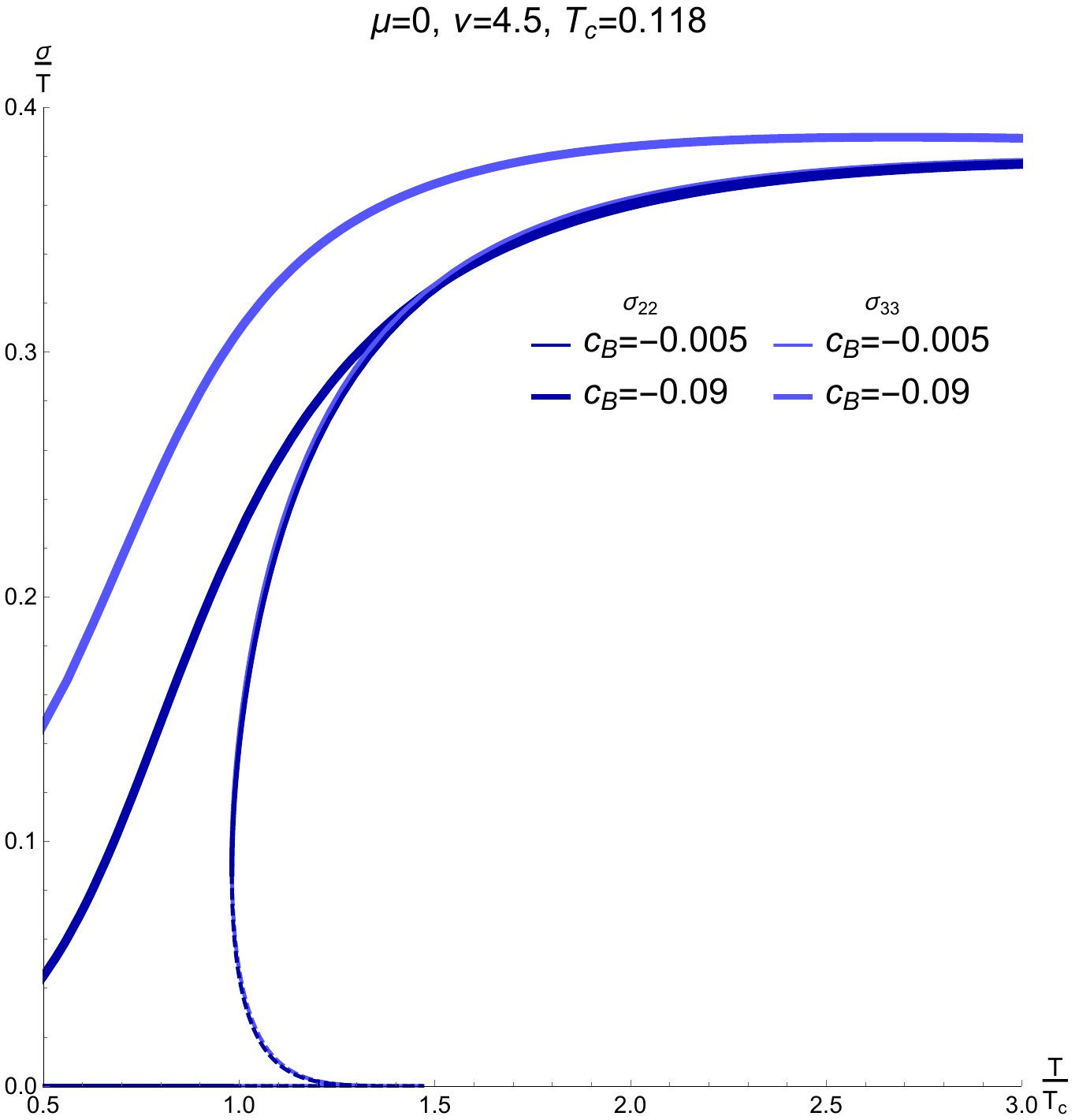} \quad \includegraphics[width=0.3\linewidth]{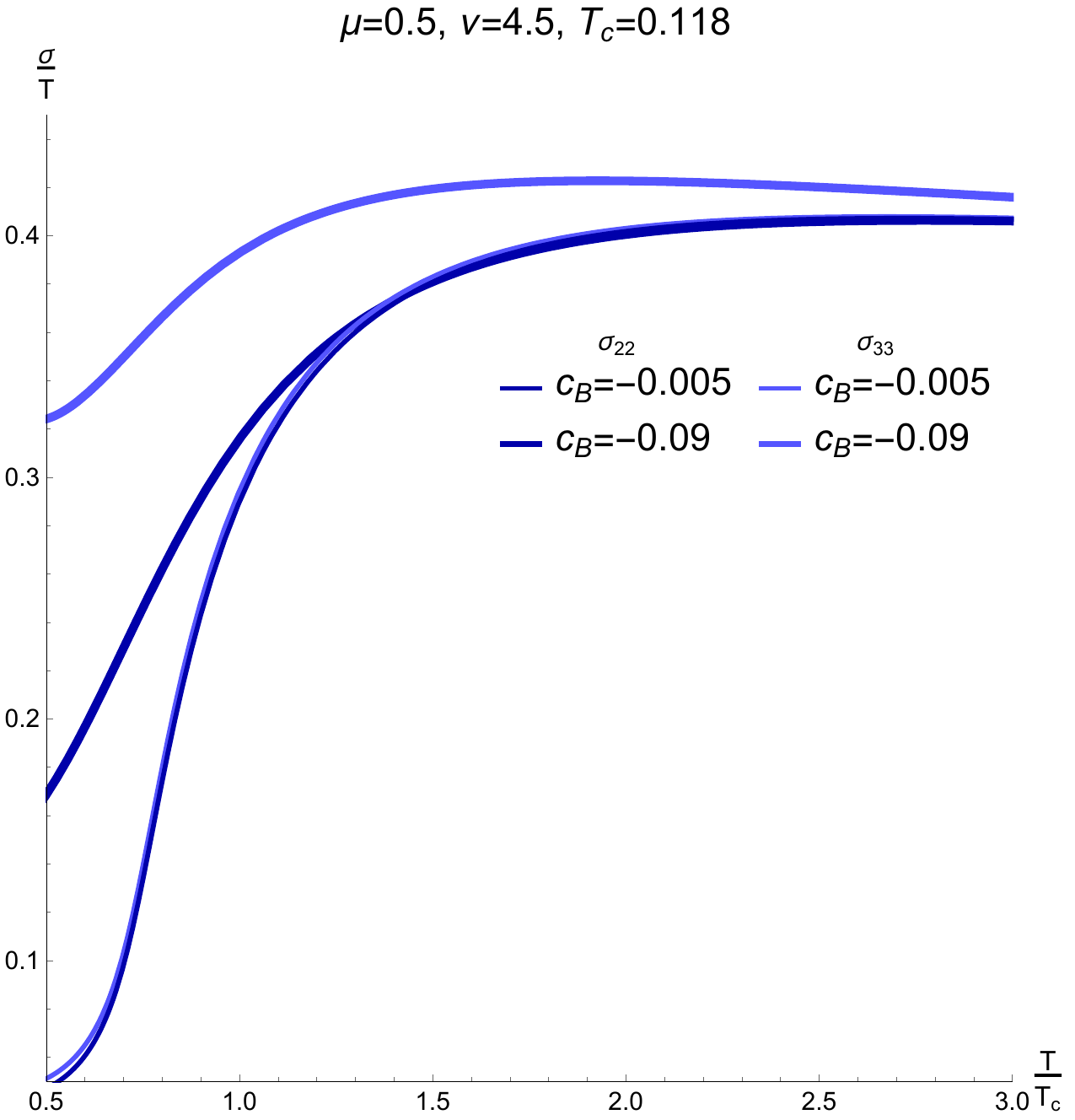}}
\\
  A \hspace{50 mm} B \hspace{50 mm} C\\
\caption{The dependence of $\sigma^{22}/T$ and $\sigma^{33}/T$ on the normalized temperature $T/T_c$ for different values of magnetic field's  parameter $c_B$ and  $\mu = 0$ (A, B), $\mu = 0.05$ (C). Here $\nu =1$ on (A), and $\nu=4.5$ on (B) and (C). Dashed lines represent values of the conductivity calculated in thermodynamically unstable phase. The inset in (A) shows  jumps near the critical points.}
\label{fig:16}
\end{figure}

It is interesting to note that $\sigma^{22}$ and $\sigma^{33}$  are not very different, see Fig.\ref{fig:16}.

We see that,  conductivities $\sigma^{22}$ and $\sigma^{33}$ are indistinguishable for small values of an external magnetic field (thin lines are almost indistinguishable in Fig.\ref{fig:16}). Neither anisotropy nor chemical potential can remove this degeneracy. It is clear from the definition of the anisotropy functions \eqref{anisfun} and the expressions for conductivities \eqref{sigma22}-\eqref{sigma33}. The anisotropy-dependant factors cancel out, and the magnetic field's parameter enters these expressions with different signs. Therefore, the magnetic field tends to push the conductivities in different directions: $\sigma^{33}$ rises, and $\sigma^{22}$ goes down. All the other thermodynamic properties of these two values are the same. 

$$\,$$

\newpage

 \section{Conclusion and Discussion} \label{ConclDiscuss}
 
We have got that formulas \eqref{sigma33}, \eqref{sigma11} and \eqref{sigma22}  give the electric conductivity for anisotropic holographic models. These formulas can be  presented in a uniform way as
\bea\label{sigma-gen}
\sigma^{ii}&=&f_0\, g^{ii}\sqrt{\det g^{(3)}_{jk}}\,\Big|_{z=z_h}, \qquad i,j,k=1,2,3,
\eea
where  $g^{(3)}_{jk}$ is the spatial part of the diagonal metric \eqref{metric-full-anis}. Noticing that the last multiplayer in \eqref{sigma-gen}  is nothing but the density of the entropy obtained in the case of the absence of the dynamical wall \cite{Arefeva:2020uec}, in the terminology of \cite{AR-2018}, we can rewrite \eqref{sigma-gen} as 
\bea\label{sigma-gen-s}
\sigma^{ij}&=&f_0\, g^{ii}\,\Big|_{z=z_h}\,s, \qquad i=1,2,3,
\eea
where $s$ is the   entropy  density.
It would be interesting to compare the dependence of the DC conductivity
\eqref{sigma33}--\eqref{sigma22} on the anisotropy functions $\mathfrak{g}_i$ 
with other characteristics of anisotropic plasma \cite{Giataganas:2012}; in particular, with the string tensions  and  the drag forces 
calculated for model \eqref{action} in \cite{Arefeva:2020vae,Arefeva:2020,Arefeva:2020bjk, Arefeva:2021btm}. It is convenient  to write these quantities in terms of vierbeins
\bea
{\bf e}^a&=&e_\mu^a(z)dx^\mu,
\quad a=0,1,...3,\quad \mu= 0,1,...3,\\
e_\mu^a(z)&=&\delta _\mu ^a\,  \sqrt{\frac{\fb}{z^2}\, \fg_{\mu}},\quad \fp_0=g,\,\, 
g_{\mu\nu}=\eta_{ab}e_\mu^a e_\nu^b,\nn\eea
$\eta_{ab}$ is the Minkowski metric. In particular, in the absence of the dynamical wall,  
  the tension of the Wilson loop lying on a (regularized) boundary and extending in the spatial directions $(i,j)$ is 
\bea
\bm{\tau }^{ab} &=&{\bf e}(z)^a\wedge {\bf e}(z)^b\Big|_{z=z_h},
\\
\tau ^{ab}_{\mu\nu}&=&e(z)^a_\mu  e(z)^b_\nu=\delta _\mu ^a
\delta _\nu ^b\frac{b}{z^2} \sqrt{\,\fp_{\mu}\,\fp_{\nu}}\Big|_{z=z_h}.
\eea
For the drag forces in anizotropic models
we have got \cite{Arefeva:2020}
\be
f_i =g_{ii}\Big|_{z=z_h}\, v^i,\quad i=1,2,3, \ee
here $v^i$ is a constant velocity.
\\

We have numerically studied the dependence of the electric conductivity on the anisotropy parameter, temperature, magnetic field, and chemical potential within the holographic anisotropic QGP model \cite{Arefeva:2020vae}. Plots of the electric conductivity vs temperature are presented in the summary Table \ref{Table:1}.
 Here the gauge kinetic function is set to 1, $f_0=1$.

We use freedom of choice of the gauge kinetic function to reproduce the lattice results for small chemical potential, magnetic field and isotropic medium. Then, we calculate the conductivity with $f_0$ from \eqref{fitfunc}.
The results are summarized in Table \ref{Table:2}. The most essential consequences are
\begin{itemize}
\item at high temperatures the ratios $\sigma^{ii}/T$, $i=2,3$ (orthogonal to the collision line directions) go to the same constant value which is defined by the choice of $f_0$ function. Meanwhile, the $\sigma^{11}/T$ goes to zero. The DC conductivity becomes so small for large $T$, that the QGP is almost opaque along the heavy-ion collision line; 
 \item near the phase transition 
 \begin{itemize} 
 \item the DC conductivity experiences a jump for small or zero values of chemical potential and magnetic field for both isotropic and anisotropic cases;
\item increasing the chemical potential and/or magnetic field smoothens the curves, and for high enough parameters the jump disappears;
\item generally, the phase transition is of the first order, but it happens to become the second order for some very special set of parameters;
\end{itemize}
\item
conductivities $\sigma^{22}$ and $\sigma^{33}$ are indistinguishable for small values of an external magnetic field. Neither anisotropy nor chemical potential can remove this degeneracy. \end{itemize}

Note that the DC conductivity characterises static low frequency fluctuations of the system.
Recently,  
 the butterfly velocity \cite{SS,MB}, that shows how fast chaotic correlations propagate in the plasma,   has been studied in holographic anisotropic models  
\cite{Gursoy:2020kjd}. 
It has been observed that it exhibits a rich structure as a function of temperature, anisotropy and magnetic field and exceeds the conformal value in certain regimes.  It would be interesting to investigate the butterfly's velocity for the model considered here \cite{Arefeva:2020vae} and for the light quarks holographic model \cite{Arefeva:2020byn}. Interplane between DC conductivity and butterfly velocity over an anisotropic background has been considered recently in \cite{Liu:2021stu}.
\\

We also plan to consider more general ansatz of a plane electromagnetic wave and to investigate the dependence of differential photon emission rate on anisotropy parameter, temperature and chemical potential, as well to use directly the holographic model for light quarks, that supposed to be a generalization for the twice anisotropic case of the model considered in \cite{Arefeva:2020byn}.

$$\,$$
\section{Acknowledgments}

We would like to thank K.Rannu for useful discussions. A.E. would like to thank A. Starinets for correspondence. This work is supported by Russian Science Foundation grant 
20-12-00200. 
$$\,$$

\newpage
\appendix
\section*{Appendix}
\section{Retarded Green's functions approach}\label{Sect:green}
To evaluate polarization operators, one has to write the on-shell action in momentum space. Therefore we Fourier transform fields as
\be
    A_\mu (z, \Vec{x},t)=\int \frac{d^4 k}{(2\pi)^4}e^{-i(wt-\Vec{k}\Vec{x})}A_\mu (z,w,k).
\ee
Also, we decompose 
\be E_i(w, k, z)=w\mathcal{E}_i(w, k)\psi_i(z),\quad  \mbox{for} \quad i=1,2,3,
\label{dec1}\ee
where $\psi_i(z)\rightarrow 1$ as $z\rightarrow 0$ \cite{Son:2002sd}. \\

The surface term \eqref{Surface} in momentum space then takes the following form
\bea
   && S_{sur\!f}=
   \int \frac{d^4 k}{(2\pi)^4}  \frac{f_0 \, g}{z} \sqrt{\fb \, \mathfrak{g}_1 \mathfrak{g}_2 \mathfrak{g}_3} \left( -\mathcal{E}_3(-k, -w) \frac{\psi_3^*\psi_3' }{\frac{k^2}{w^2}g-\mathfrak{g}_3} \mathcal{E}_3(k, w) + \right.  \nn \label{on-shell}\\
    &&\left.\left. +\mathcal{E}_1(-k, -w)  \frac{\psi_1^* \psi_1'}{\mathfrak{g}_1} \mathcal{E}_1(k, w)+\mathcal{E}_2(-k, -w)\frac{\psi_2^* \psi_2'}{\mathfrak{g}_2}\mathcal{E}_2(k,w) \right) \right| _{z=0}^{z=z_h}.
\eea
2-point Green's functions are defined from the action \eqref{on-shell} as
\be
    G^{\mu \nu}(k_1,\omega_1;k_2,\omega_2)=\frac{\delta^2 S_{sur\!f}}{\delta \mathcal{E}_{\mu}(k_1,\omega_1)  \delta \mathcal{E}_{\nu}(k_2,\omega_2)} .
\ee
To calculate Green's functions one has to know functions $\psi$ on both the horizon and the boundary. These functions satisfy equations of motion for Maxwell field in the bulk. Furthermore, one needs physically reasonable boundary conditions to fix these solutions. To overcome these obstacles, we are following the prescription from \cite{Son:2002sd}. According to it, one has to find asymptotic solutions to EOMs which are constant on the boundary and satisfy the in-falling conditions on the horizon. One then obtains the on-shell action in the form 
\begin{equation}
    S=\int \left.  \frac{d^4 k}{(2 \pi )^4} J(-k) \mathcal{F}(z,k) J(k) \right| _{z=0}^{z=z_h} .
\end{equation}
Following \cite{Policastro} the retarded Green's function is
\begin{equation}
    G_R(k)=-2\lim _{z\rightarrow 0}\mathcal{F}(z,k), \label{Retard}
\end{equation}
and its imaginary part as in \cite{Son:2002sd} can be represented as
\be
    \mbox{Im} G_R(k)=-2\lim _{z\rightarrow z_h}\mbox{Im} \mathcal{F}(z,k). 
\ee
To calculate retarded Green's functions we plug the decomposition \eqref{dec1} into \eqref{eq: Eel}-\eqref{eq: Ee2} and investigate asymptotics at $z=0$ and $z=z_h$. 

We assume that the asymptotic value of the blackening function on the boundary is
\begin{equation}
    g(z)= 1 + o(z); \label{gonbdr}
\end{equation}
while on the horizon is
\begin{equation}
    g(z)=g'(z_h)(z_h-z)+o(z_h-z). \label{gonhor}
\end{equation}

\subsection{Transverse components of electric field}
First, we find asymptotics for \eqref{eq: Ee1}
\bea
    &\psi_1'' + \psi_1' \left(\dfrac{\fb'}{2 \fb}+\dfrac{f_0'}{f_0}+\dfrac{g'}{g}+\dfrac{\mathfrak{g}_2'}{2 \mathfrak{g}_2}+\dfrac{\mathfrak{g}_3'}{2 \mathfrak{g}_3}-\dfrac{\mathfrak{g}_1'}{2 \mathfrak{g}_1}-\dfrac{1}{z}\right)+\psi_1\dfrac{w^2 \mathfrak{g}_3-k^2 g}{g^2 \mathfrak{g}_3}=0. \label{psi1}
\eea
The requirement $\psi_i \rightarrow 1$ near the boundary is consistent with equations of motion. Its' asymptotic behaviour near the boundary $z=0$ takes the form
\begin{equation}
    \psi_1'' -\frac{ \psi_1'}{z}=0. \label{as1}
\end{equation}
Its solution is simply
\begin{equation}
    \psi_1=C_1+C_2 z^2 .
\end{equation}
\\
Near the horizon \eqref{psi1} has singularities of $1/g$ terms. So the equation for asymptotic behaviour on the horizon 
\begin{equation}
    \psi_1'' -\frac{1}{z_h-z} \psi_1'+\frac{w^2}{g'(z_h)^2 (z_h-z)^2}\psi_1=0.
\end{equation}
The solution for this equation is a linear combination of two linear independent solutions
\begin{align}
    \psi _1(z)=&c_1\cos\left( \frac{w}{g'(z_h)}\ln(z_h-z)\right) +c_2\sin\left( \frac{w}{g'(z_h)}\ln(z_h-z)\right) =\\\nonumber
    & = \frac{1}{2} (c_1+i c_2) (z_h-z)^{-i\frac{w}{g'(z_h)}}+\frac{1}{2} (c_1-i c_2) (z_h-z)^{i\frac{w}{g'(z_h)}}.
\end{align}
Both these solutions oscillate and are finite near the horizon. We will call the first one $f_{k^{\mu}}=(z_h-z)^{-i\frac{w}{g'(z_h)}}$ and the second one as its complex conjugate $f_{-k^{\mu}}$. If one also restores time-dependent part $e^{-iwt}$, then 
\begin{equation}
e^{-iwt}f_{k^{\mu}}=e^{-iw(t+z_*)}, 
\end{equation}
where $z_*=\frac{\ln (z_h-z)}{g'(z_h)}$. Therefore $f_{k^{\mu}}$ describes the incoming (that moves towards the horizon) wave, whilst $f_{-k^{\mu}}$ corresponds to outgoing wave. The boundary condition on the horizon requires choosing the incoming wave only since no signal can escape from a black hole. So we have \begin{equation}
    \mathcal{F}(z,k)=\frac{f_0 g}{z}\sqrt{\frac{\fb \,\mathfrak{g}_3 \mathfrak{g}_2}{\mathfrak{g}_1}}f_{-k^{\mu}}\frac{d f_{k^{\mu}}}{dz}=\frac{f_0 g}{z}\sqrt{\frac{\fb \,\mathfrak{g}_3 \mathfrak{g}_2}{\mathfrak{g}_1}}\frac{iw}{g'(z_h)}\frac{1}{z_h-z}. \label{F}
\end{equation}
\bea
    \mbox{Im}\, G_R^{11}&=& -2 \lim _{z\rightarrow z_h} \frac{f_0(z) g(z)}{z}\sqrt{\frac{\fb(z) \,\mathfrak{g}_3(z) \mathfrak{g}_2(z)}{\mathfrak{g}_1(z)}}\frac{w}{g'(z_h)(z_h-z)}\nn\\ &=&
-2 w f_0(z_h)\sqrt{\frac{\fb(z_h)\mathfrak{g}_3(z_h) \mathfrak{g}_2(z_h)}{\mathfrak{g}_1(z_h) z^2_h}}.\label{ImGR}
\eea

And using the Kubo formula $\sigma^{\mu \nu} = -G^{\mu \nu}_R/iw$ we obtain the  11-component of QGP electric  conductivity tensor to be nothing but \eqref{sigma11}.
Doing all the same we obtain the 22-component of QGP electric  conductivity to be \eqref{sigma22}.
\subsection{Longitudinal component of electric field}
From the on-shell action \eqref{on-shell} the longitudinal part reads
\begin{equation}
    S_{sur\!f}=\int \frac{d^4 k}{(2 \pi)^4}f_0\frac{g}{z}\mathcal{E}_3(-k,-w)\psi_3^*\psi_3'\mathcal{E}_3(k,w)\frac{\sqrt{\fb\mathfrak{g}_3 \mathfrak{g}_2 \mathfrak{g}_1}}{\mathfrak{g}_3-g\frac{k^2}{w^2}   },
\end{equation}
and the equation of motion \eqref{eq: Ee}
\bea
    &\psi_3'' + \psi_3' \left(\dfrac{\fb'}{2 \fb}+\dfrac{f_0'}{f_0}-\dfrac{w^2 \mathfrak{g}_3 g'}{k^2 g^2-w^2 g \mathfrak{g}_3}+\dfrac{w^2 \mathfrak{g}_3'}{k^2 g-w^2 \mathfrak{g}_3}+\dfrac{\mathfrak{g}_1'}{2 \mathfrak{g}_1}+\dfrac{\mathfrak{g}_2'}{2 \mathfrak{g}_2}+\dfrac{\mathfrak{g}_3'}{2 \mathfrak{g}_3}-\dfrac{1}{z}\right)+\nn\\
    &+\psi_3\dfrac{w^2 \mathfrak{g}_3-k^2 g}{g^2 \mathfrak{g}_3}=0.
\eea
To find the asymptotic behaviour of the solution to this equation near the horizon, we make the same assumptions about functions as in the previous section \eqref{gonbdr}, \eqref{gonhor}. Then the equation in the vicinity of horizon is

\begin{equation}
    \psi_3''+\frac{\psi_3'}{(z-z_h)}+\frac{w^2}{(g'(z_h))^2(z-z_h)^2}\psi_3=0.
\end{equation}
Its' solution is 
\begin{equation}
    \psi_3(z)=c_1(z_h-z)^{-\frac{iw}{g'(z_h)}}+c_2(z_h-z)^{\frac{iw}{g'(z_h)}}.
\end{equation}
Where we as before imply the infalling boundary condition, eliminating the outgoing wave.
\be
    \psi_3(z) \sim (z_h-z)^{-iw/g'(z_h)}.
\ee
Then the imaginary part of the retarded Green's function is
\begin{equation}
    \mathrm{Im}G_R^{33}=-2\lim _{z\rightarrow z_h}\mathrm{Im}\left[ \frac{f_0 g}{z}\frac{\sqrt{b \, \mathfrak{g}_1 \mathfrak{g}_2 \mathfrak{g}_3}}{\mathfrak{g}_3-g\frac{k^2}{w^2}}\psi_3^* \frac{d\psi_3}{dz} \right]
\end{equation}
In the $z \rightarrow z_h$ limit we use the obtained asymptotic behaviour of $\psi_3$
\begin{equation}
    \psi_3^* \frac{d\psi_3}{dz}=\frac{iw}{g'(z_h)(z_h-z)}.
\end{equation}

The result then follows from the Kubo relation \eqref{sigma33}. \\
Note, all the obtained results completely agree with another approach from Sect. \ref{Sect:conduct} and reduce to that of isotropic and partially anisotropic models. 

\newpage
\section{Tables for conductivities}\label{Sect:Tables}
\begin{table}[h]
\center{\includegraphics[width=\linewidth]{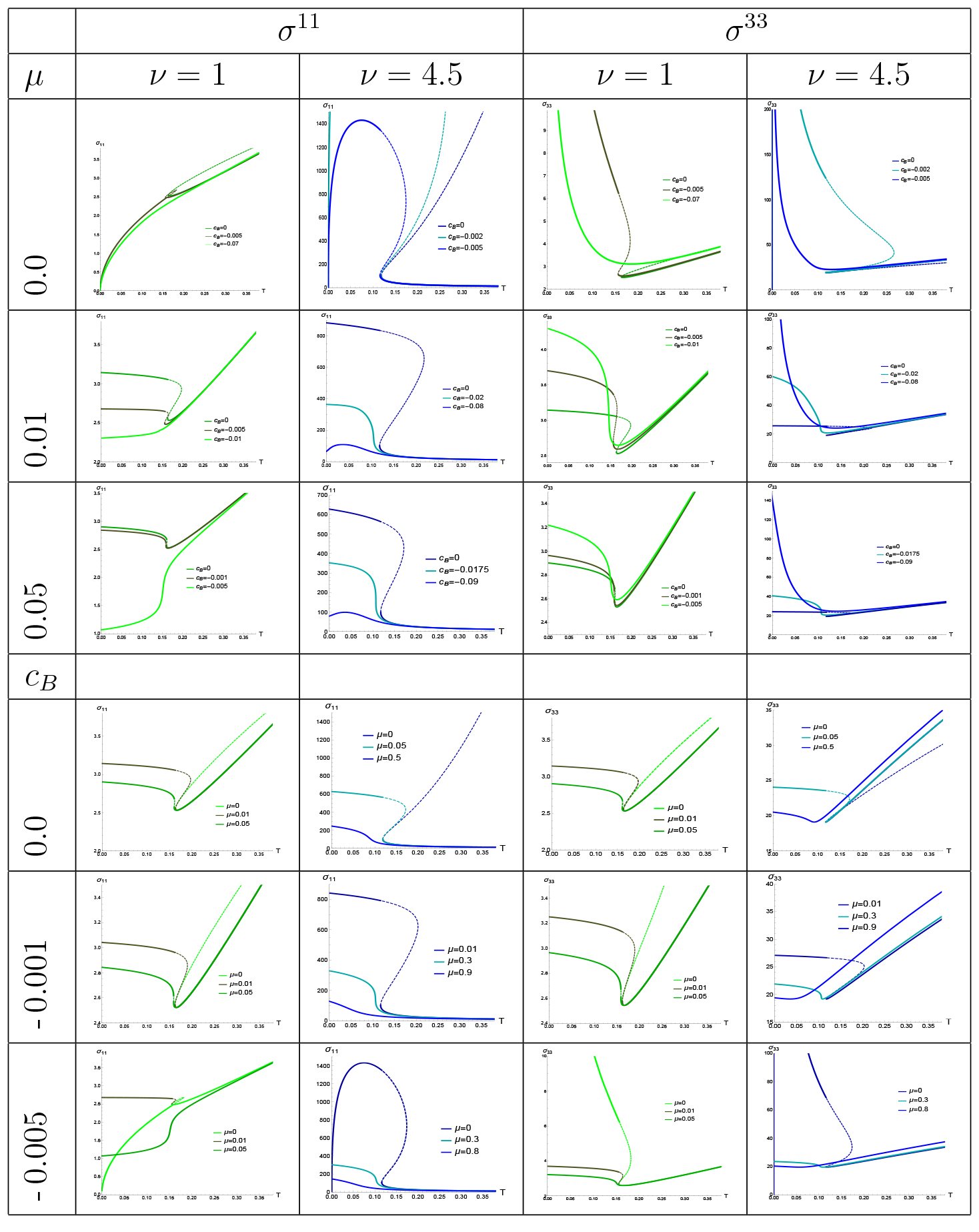}}
\caption{Table shows the dependence of the DC conductivity 
on anisotropy, chemical potential and magnetic field. Here $f_0=1$.}\label{Table:1}
\end{table}

\begin{table}[h]
\center{\includegraphics[width=\linewidth]{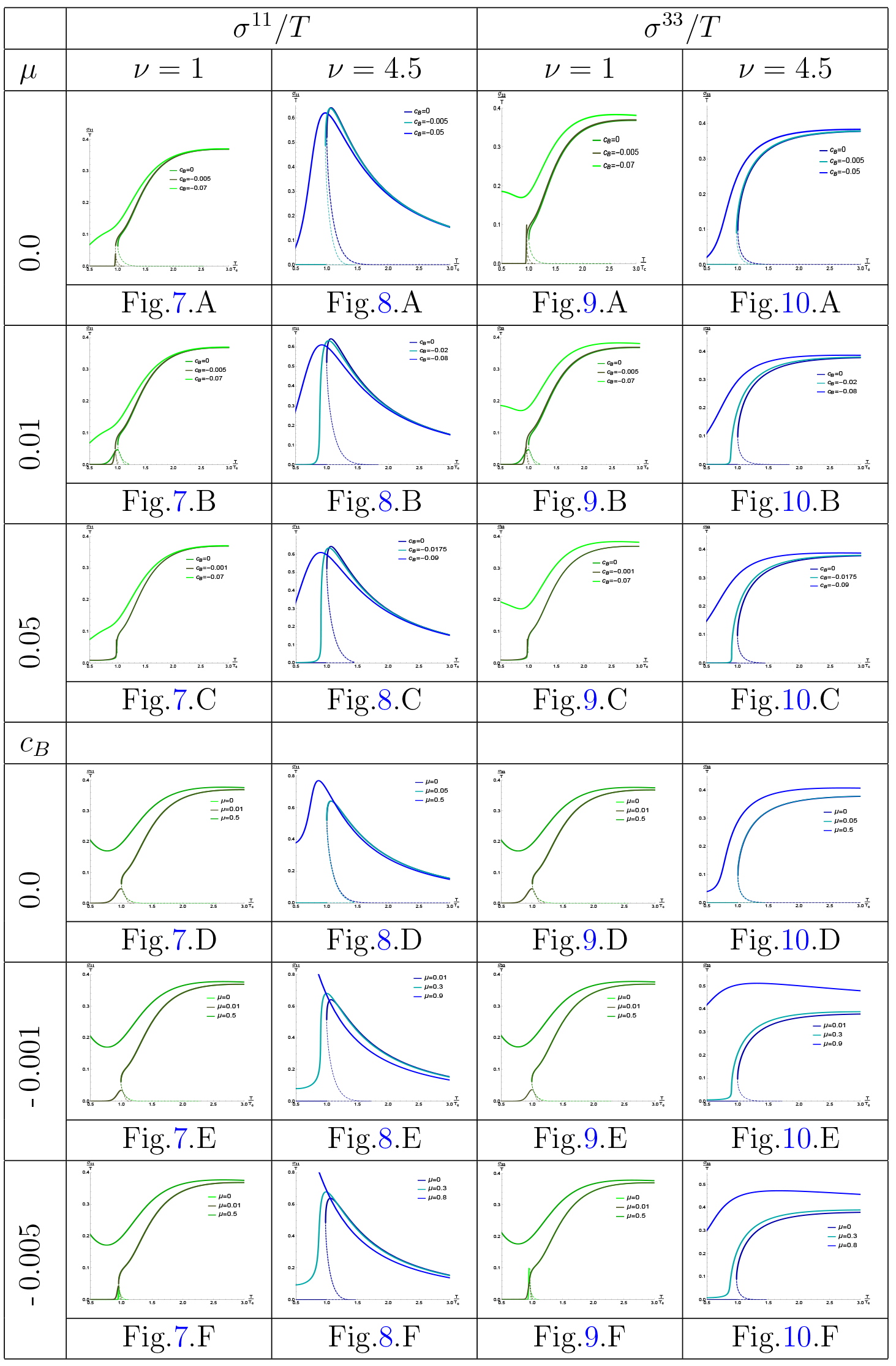}}
\caption{Table shows dependencies of the ration of the DC conductivity to temperature 
on anisotropy, chemical potential and magnetic field. Here $f_0$ is given by \eqref{fitfunc}}\label{Table:2}
\end{table}
\clearpage






\end{document}